\documentclass[12pt]{article}
\usepackage{amssymb,amsmath}
\usepackage{epsfig,psfrag}
\usepackage{curves}

\catcode`\@=11
\def \Tr {\mathop{\rm Tr}\nolimits}
\def \tr {\mathop{\rm tr}\nolimits}

\newcommand\lr[1]{{\left({#1}\right)}}

\newcommand\re[1]{(\ref{#1})}
\def \qqquad {\qquad\quad}
\def \qqqquad {\qquad\qquad}

\newcommand{\dslash}{\partial\hskip-6.28pt /}
\newcommand{\pa}{\partial}

\newcommand{\nn}{\nonumber}

\renewcommand{\a}{\alpha}

\newcommand{\da}{{\dot\alpha}}

\renewcommand{\b}{\beta}

\newcommand{\q}{\theta}
\newcommand{\bq}{\bar\theta}

\newcommand{\ep}{\epsilon}

\newcommand{\cN}{{\cal N}}

\newcommand{\rmd}{{\rm d}}
\newcommand{\p}[1]{(\ref{#1})}
\newcommand{\bt}[1]{{\bar t}}
\newcommand{\ts}{\textstyle}

\newcommand{\half}{{\ts \frac{1}{2}}}
\newcommand \vev [1] {\langle{#1}\rangle}

\newcommand {\cO}{{{\cal O}}}
\newcommand {\tO}{{{\tilde{\cal O}}}}
\newcommand{\beq}{\begin{equation}}
\newcommand{\eeq}{\end{equation}}
\newcommand{\be}{\begin{equation}}
\newcommand{\ee}{\end{equation}}
\newcommand{\bea}{\begin{eqnarray}}
\newcommand{\eea}{\end{eqnarray}}
\newcommand{\ena}{\end{eqnarray}}
\newcommand{\bear}{\begin{eqnarray}}
\newcommand{\ear}{\end{eqnarray}\noindent}

\newcommand{\ft}[2]{{\textstyle\frac{#1}{#2}}}

\def\numberbysection{\@addtoreset{equation}{section}
                     \def\theequation{\thesection.\arabic{equation}}}
\numberbysection

\begin{document}


\thispagestyle{empty}

\null\vskip-12pt \hfill
\begin{minipage}[t]{35mm}
 LU-ITP 2010/003 \\
  IPhT--T10/92 \\
 LAPTH--026/10
\end{minipage}

\vskip3.2truecm
\begin{center}
\vskip 0.2truecm {\Large\bf
From correlation functions to  scattering amplitudes}

\vskip 1truecm

{\bf    Burkhard Eden$^{*}$, Gregory P. Korchemsky$^{\dagger}$, Emery Sokatchev$^{\ddagger}$ \\
}

\vskip 0.4truecm
$^{*}$ {\it  Institut f\"ur theoretische Physik, Universit\"at Leipzig \\
Postfach 100920, D-04009 Leipzig, Germany \\
 \vskip .2truecm
$^{\dagger}$ Institut de Physique Th\'eorique\,\footnote{Unit\'e de Recherche Associ\'ee au CNRS URA 2306},
CEA Saclay, \\
91191 Gif-sur-Yvette Cedex, France\\
\vskip .2truecm $^{\ddagger}$ LAPTH\,\footnote[2]{Laboratoire d'Annecy-le-Vieux de Physique Th\'{e}orique, UMR 5108},   Universit\'{e} de Savoie, CNRS, \\
B.P. 110,  F-74941 Annecy-le-Vieux, France
                       } \\
\end{center}

\vskip 1truecm 
\centerline{\bf Abstract} 
\medskip
\noindent

We study the correlators of half-BPS protected operators in $\cN=4$ super-Yang-Mills theory, in the limit where the positions of the adjacent operators become light-like separated. We compute the loop corrections by means of Lagrangian insertions. The divergences resulting from the light-cone limit are regularized by changing the dimension of the integration measure over the insertion points. Switching from coordinates to dual momenta, we show that the logarithm of the correlator is identical with twice the logarithm of the matching MHV gluon scattering amplitude. We present a number of examples of this new relation, at one and two loops.

\newpage

\thispagestyle{empty}

{\small \tableofcontents}

\newpage
\setcounter{page}{1}\setcounter{footnote}{0}



\section{Introduction}

One of the most remarkable manifestations of the AdS/CFT correspondence \cite{Maldacena:1997re} in the recent years was the duality between planar gluon scattering amplitudes and light-like polygonal Wilson loops. It was first proposed at strong coupling \cite{am07} and soon afterwards also observed at weak coupling, first at one loop  \cite{Drummond:2007aua,Brandhuber:2007yx}, followed by extensive two-loop tests \cite{Drummond:2007cf,Drummond:2007au,Drummond:2008aq,Bern:2008ap}. This duality can be formulated as follows:
\begin{equation}\label{formdu}
    \ln\left(A_n/A^{(0)}_n\right) = \ln\left(W[C_n]\right) + O(1/N_c) +O(\ep)\,.
\end{equation}
Here $A_n$ is the all-order $n-$gluon MHV scattering amplitude depending on the particle light-like four-momenta $p_i$ (with $p^2_i=0$ and $\sum_{i=1}^n p_i=0$), and $A^{(0)}_n$ is the tree-level amplitude. An essential step in establishing the relation \p{formdu} is the so-called T-duality transformation from  momenta to dual coordinates:
\begin{equation}\label{moco}
    p_i = x_i - x_{i+1} \equiv x_{i,i+1}\,, \qquad x^2_{i,i+1}=0\,, \qquad x_{i+n} \equiv x_i\,.
\end{equation}
The Wilson loop $W[C_n]$ is defined on a closed polygonal contour $C_n$ in the dual space, with cusps at points $x_i$ and with light-like sides $[x_i, x_{i+1}]$. The duality \p{formdu} holds in planar $\mathcal{N}=4$ super-Yang-Mills theory (SYM) and up to terms vanishing if the suitably identified infrared (for amplitudes) and ultraviolet (for Wilson loops) regulators $\ep\to 0$.\footnote{Quite remarkably, the duality between planar amplitudes and light-like Wilson loops also holds in gauge theories with less or no supersymmetry, including QCD. However, in distinction with $\mathcal{N}=4$ SYM, there the
relation \re{formdu} is satisfied in the high-energy (Regge) limit only~\cite{Korchemskaya:1996je,Drummond:2007aua}.}

An important ingredient in this duality is the notion of dual conformal symmetry. This is the natural symmetry of the light-like Wilson loop, becoming anomalous due to the cusp singularities~\cite{P80,KR87,KK92}. By virtue of the duality \p{formdu} it is extended to a symmetry of the planar scattering amplitudes of {\it dynamical} origin. The first evidence for this new symmetry came from the study \cite{Drummond:2006rz} of the loop momentum integrals appearing in the four-gluon amplitudes up to four (or even five) loops \cite{Anastasiou:2003kj,Bern:2005iz,Bern:2006ew,Bern:2007ct}. Once rewritten in dual space according to \p{moco}, they become pseudo-conformal (the infrared regulator breaks the symmetry). This dual conformal symmetry, or rather its anomalous version \cite{Drummond:2007au,Drummond:2007bm}, was instrumental in explaining the so-called BDS ansatz for MHV amplitudes \cite{Bern:2005iz}.

{In the present paper we provide evidence for another duality relation in the $\cN=4$ SYM theory, this time between the MHV gluon amplitudes and the correlation functions of gauge invariant composite operators on the light cone. The operators we consider belong to the class of half-BPS (or ``short") scalar operators. They are of the type $\cO^{(k)} = \Tr(\phi^k)$, made of the six real scalars $\phi$ of the $\cN=4$ SYM theory. They carry R-symmetry $SU(4)$ Dynkin labels $[0,k,0]$ and transform as chiral primaries under the superconformal symmetry $PSU(2,2|4)$ of the $\cN=4$ theory, with fixed conformal dimension $d=k$. In perturbation theory, such operators do not undergo renormalization and are thus protected to all orders. The best known example is the simplest, bilinear ($k=2$) operator, belonging to the so-called stress-tensor superconformal multiplet. In the context of the AdS/CFT correspondence, these operators are dual to massive Kaluza-Klein modes in the compactification of type IIB supergravity on an AdS${}_5 \times S^5$ background.

The correlators of half-BPS operators have been the subject of numerous studies. Not only the conformal dimension of the operators, but also their two- and three-point correlation functions are protected \cite{Penati:1999ba,Penati:2000zv,D'Hoker:1998tz,Lee:1998bxa,Howe:1998zi}. The first non-trivial quantum corrections appear in the four-point correlators of protected bilinear operators, which have been computed up to two loops in \cite{Eden:1998hh,GonzalezRey:1998tk,Eden:1999kh,Eden:2000mv,Bianchi:2000hn}. The knowledge of these quantum corrections allowed one to extract the spectrum of anomalous dimensions of the Konishi operator \cite{Bianchi:2001cm}, and later on of all twist-two operators up to two loops by means of a conformal operator product expansion (OPE) \cite{Dolan:2000ut}.

Here we propose to look at such correlators from a novel point of view. Consider the correlation function of $n$ protected operators
\begin{equation}\label{nO}
   G_n= \vev{\cO(x_1)\cO(x_2)\dots \cO(x_n)}\,.
\end{equation}
As long as we maintain the points $x_i$ (with $i=1,\ldots,n$) in generic positions, this function is well defined and has conformal symmetry. {As a consequence, it is given
by a product of free scalar propagators times some (coupling dependent) function of conformal cross-ratios $x_{ij}^2 x_{kl}^2/(x_{ik}^2x_{jl}^2)$. In perturbation theory this function is expressed in terms of conformally invariant space-time loop integrals.}
Now, imagine that we wish to take the limit in which the neighboring points become light-like separated,\footnote{A similar light-cone limit has extensively been studied in QCD, see, e.g., the review \cite{BKM}.}
\begin{equation}\label{llse}
    x^2_{i,i+1} \to 0\,, \qquad x_{i+n} \equiv x_i\,.
\end{equation}
The correlator $G_n$ becomes singular in this limit.
The first problem we have to face are the pole singularities in $G^{(0)}_n$, due to the propagators $1/x^2_{i,i+1}$ connecting two neighboring scalars. Secondly, the loop integrals develop logarithmic light-cone divergences $\sim \ln x_{i,i+1}^2$ when the integration points approach one of the light-like segments $[x_i, x_{i+1}]$. To deal with the first problem, it is sufficient to consider the ratio $G_n/G^{(0)}_n$, in which the pole singularities are removed. The second problem is more serious, it requires introducing an appropriate regularization.

Two possible choices of a regularization procedure {are:  (i) use the small distances $x^2_{i,i+1}$ as a cutoff;  (ii) employ standard dimensional regularization
and set $x^2_{i,i+1}=0$ from the very beginning.} {These two regularizations
are  considered in the parallel publication \cite{AEKMS}, where it is shown that
in both cases the correlation function reduces to a Wilson loop,
\begin{equation}\label{cowll}
    \lim_{x^2_{i,i+1} \to 0} G_n/G^{(0)}_n  \propto \left(W[C_n]\right)^2\,.
\end{equation}
The exact form of the proportionality factor in the right-hand side of this relation depends on the regularization; for case (ii) it is just $1$.} Here $W[C_n]$ is the light-like polygonal Wilson loop in the fundamental representation of the gauge group described earlier. Since $W[C_n]$ is dual to the MHV gluon amplitude, Eq.~\re{formdu}, {we expect that the
ratio of the correlation functions in the left-hand side of \re{cowll} is also related to the ratio of amplitudes, $A_n/A_n^{(0)}$. The question arises if we could find another, more direct way of establishing the relation between correlation functions and amplitudes without invoking Wilson loops. }

In the present paper we propose a scenario which realizes this direct link. It employs an unusual, {\it dual infrared} dimensional regularization procedure. This may seem surprising, as we have just argued that the singularities of the correlator occur at short distances. Nevertheless, we can consider the following alternative. We start by computing the loop corrections to the correlator by means of Lagrangian insertions. This method is well known in field theory and has been further developed in \cite{Eden:1999kw,Eden:2000mv} for the two-loop computations of four-point correlators, but it has universal applicability. The idea is to interpret the loop corrections to the correlator as derivatives with respect to the coupling $g$. For instance, the one-loop correction
\begin{equation}
   g^2 \frac{\pa}{\pa g^2} G_n = -i\int d^D x_0 \, {\cal G}^{(0)}_{n+1}(x_0; x_1,  \ldots, x_n) \label{g2'}
\end{equation}
is calculated from the {\it Born-level}  $(n+1)$-point correlator
\begin{equation}\label{g41'}
    {\cal G}^{(0)}_{n+1}(x_0; x_1,  \ldots, x_n) = \vev{ L(x_0) \cO(x_1)\ldots {\cal O}(x_n)}
\end{equation}
obtained by inserting the Lagrangian at the extra point $x_0$. The crucial point here is that the correlator \p{g41'} stays well defined (after dividing it by the tree approximation $G^{(0)}_n$) in $D=4$ dimensions, even if we put the outer points $x_i$ on the light cone, but keeping the insertion point $x_0$ in a generic position. The logarithmic singularities originate from the integration over the insertion point in \p{g2'}. Then we propose to regularize this integral by choosing a measure in $D=4-2\ep$ dimensions, with $\ep<0$. This unusual regularization is motivated by the analogy between the space-time loop integrals appearing in \p{g2'}, and the momentum loop integrals in the gluon MHV scattering amplitude $A_n$ discussed earlier. The analogy becomes possible after the T-duality transformation (or change of variables in the integrals) \p{moco}, provided we use the  infrared-like regulator above. Still, nothing guarantees at this stage that we will find a result, not only similar, but identical with a scattering amplitude. Yet, rather surprisingly, this is what happens.

In the present paper we show a number of examples of this new phenomenon. These include all the $n-$points correlators at one loop, and the four- and five-point correlator up to two loops. At present we have no explanation why this is so, but if this new duality correlators/amplitudes is confirmed, it will provide the natural explanation of the mysterious dual conformal symmetry \cite{Drummond:2006rz} of the loop momentum integrals in  all available gluon amplitude calculations ($n$ gluons at one loop \cite{Bern:1994zx}, four gluons up to five loops \cite{Anastasiou:2003kj,Bern:2005iz,Bern:2006ew,Bern:2007ct}, five \cite{Bern:2006vw} and six  \cite{Bern:2008ap} gluons up  to two loops).

The paper is organized as follows. In Sect.~\ref{pon4sym} we show a simple example of a correlation function becoming an amplitude in the light-cone limit and we formulate the main idea of the relation correlators/amplitudes. In Sect.~\ref{Cpos} we give a more detailed description of the correlators of protected half-BPS operators in superspace. We explain how their loop corrections can be obtained by Lagrangian insertions. We then outline the procedure of establishing the duality with amplitudes, in particular the introduction of dual infrared regularization.  In Sect.~\ref{n1l} we show in detail how the duality works for $n-$point correlators of bilinear half-BPS operators at one loop. In Sect.~\ref{fpcfga} this is expended to four-point correlators up to two loops, and in Sect.~\ref{52loop}  to five-point correlators up to two loops.
Appendix A summarizes some key points of the harmonic superspace formalism which we employ for loop calculations. In Appendix B we generalize our findings to four-point correlators of half-BPS operators of arbitrary weight.

\section{A simple example of the duality correlators/amplitudes}\label{pon4sym}

{In this section we discuss some general properties of the protected operators in $\cN=4$ SYM and of their correlation functions. To illustrate our main idea, we show how the four-point one-loop correlator is transformed into the four-gluon one-loop amplitude in the light-cone limit. 

The gauge invariant composite operators in $\cN=4$ SYM can be classified  as representations of the superconformal group $PSU(2,2/4)$   (see, e.g., \cite{Andrianopoli:1998jh} and references therein). There are two basic types of such operators, usually referred to as protected (or  ``short") and unprotected (or ``long"). The former satisfy conditions of BPS shortening, i.e. they are annihilated by a fraction of the supercharges. This, together with the conditions that they are superconformal primaries, implies that they have quantized, or protected conformal dimension equal to their canonical dimension. The long operators correspond to generic superconformal representations, they receive quantum corrections and acquire anomalous dimensions.

In this paper we will consider only operators of the half-BPS type. Their lowest components (or superconformal primaries) are made of the six real scalars in the $\cN=4$ vector multiplet, $\phi_{AB} = -\phi_{BA}  = \ft12  \ep_{ABCD}\bar\phi^{CD}$, where $A,B=1,\ldots,4$ are indices of the fundamental irrep of the R symmetry group $SU(4)$. Generically, they are of the type $\cO^{(k)} = \Tr(\phi^k)$, carry $SU(4)$ Dynkin labels $[0,k,0]$ and have fixed conformal dimension $d=k$. The best known example is the simplest, bilinear ($k=2$) operator, belonging to the so-called stress-tensor superconformal multiplet. The top spin state in this multiplet is the stress tensor, while the state of highest dimension is the Lagrangian of the $\cN=4$ SYM theory. The lowest dimension state of the multiplet is the bilinear scalar operator
\begin{equation}\label{bil}
    \cO_{ABCD} = {\rm Tr}(\phi_{AB} \phi_{CD}) -  \frac1{12}\ep_{ABCD} {\rm Tr}(\bar\phi^{EF} \phi_{EF})
\end{equation}
belonging to the irrep ${\bf 20'} =[020]$ of $SU(4)$. {Here $\phi_{AB} = \phi_{AB}^a t^a$, where $t^a$ are the generators of the fundamental representation of the gauge group $SU(N_c)$, normalized as $\tr (t^a t^b) =\ft12 \delta^{ab}$. In what follows we always assume the planar limit,
\begin{align}\label{thooft}
a=\frac{g^2N_c}{8\pi^2}\,,\qqquad N_c \to \infty\,.
\end{align}
} }

Let us consider certain projections  of \re{bil}, namely
\begin{align}\label{o2}
\cO = {\rm Tr}(\phi_{12} \phi_{12})\,,\qquad
\tilde\cO = {\rm Tr}(\bar\phi^{12} \bar\phi^{12})\,,\qquad
\hat\cO= 2 \, {\rm Tr}(\bar\phi^{12} \phi_{12}) -  \frac{1}{6}
{\rm Tr}(\bar\phi^{EF} \phi_{EF})\,,
\end{align}
where $\cO$ is the (complex) highest-weight state, $\tilde\cO$ is the conjugate lowest-weight state and  $\hat\cO$ is a real projection.
We want to evaluate the correlator of $n$ such operators.  For $n=2m$ we can take, e.g.,  $m$ operators $\cO$ and $m$ conjugates $\tO$ and consider the correlator
\begin{equation}\label{defco'}
    G_n = \vev{\cO(x_1) \tilde\cO(x_2)\ldots \cO(x_{n-1})\tilde\cO(x_n)}\,.
\end{equation}
For $n=2m+1$ we can add one operator $\hat\cO$, replacing \p{defco'} by
\begin{equation}\label{defco''}
    G_n = \vev{\cO(x_1) \tilde\cO(x_2)\ldots \cO(x_{n-2})\tilde\cO(x_{n-1}) \hat\cO(x_n)}\,.
\end{equation}

Such correlators are finite (the operators \re{o2} are not renormalized) and conformally covariant, as long as the points $x_i$ are kept apart, $x_i \neq x_j$. If we let two points get close to each other, $x_i \to x_j$, we are dealing with the well-known  {short distance} expansion of the product of operators $\cO(x_i) \cO(x_j)$ mentioned above. Here we plan to do something else. We wish to take the limit where the neighboring points become light-like separated,  without coinciding with each other\footnote{The standard OPE is done in the Euclidean regime, where $x^2_{i,i+1}=0$ implies $x_i = x_{i+1}$. The Minkowski regime allows us to consider the new possibility \p{lim'}.}
\begin{equation}\label{lim'}
    x^2_{i,i+1} \ \to \ 0\,,  \qquad x_i \neq x_{i+1}\,, \qquad (i=1,\ldots,n)\,,
\end{equation}
(with the cyclic condition $x_{n+1} \equiv x_1$). This limit is singular for two reasons. Firstly, the correlator develops pole singularities, as can be seen already from the (connected, planar) tree-level approximation
\begin{equation}\label{tre''}
    G_n^{(0)} =   \frac{(2\pi)^{-2n}N^2_c}{x^2_{12} x^2_{23} \ldots x^2_{n1}} + \mbox{subleading terms}\,.
\end{equation}
By ``subleading" we mean terms corresponding to different Wick contractions of the scalar fields $\phi$ which are less singular in the limit \p{lim'} (see an illustration in Fig.~\ref{ttree}).
\begin{figure}[h!]
\psfrag{dots}[cc][cc]{$\mathbf{\dots}$}
\psfrag{1}[cc][cc]{$x_1$}
\psfrag{2}[cc][cc]{$x_2$}
\psfrag{3}[cc][cc]{$x_3$}
\psfrag{4}[cc][cc]{$x_4$}
\psfrag{n-1}[cc][cc]{$x_{n-1}$}
\psfrag{n}[cc][cc]{$x_n$}
\psfrag{(a)}[cc][cc]{(a)}
\psfrag{(b)}[cc][cc]{(b)}
%
\centerline{ \includegraphics[height=50mm]{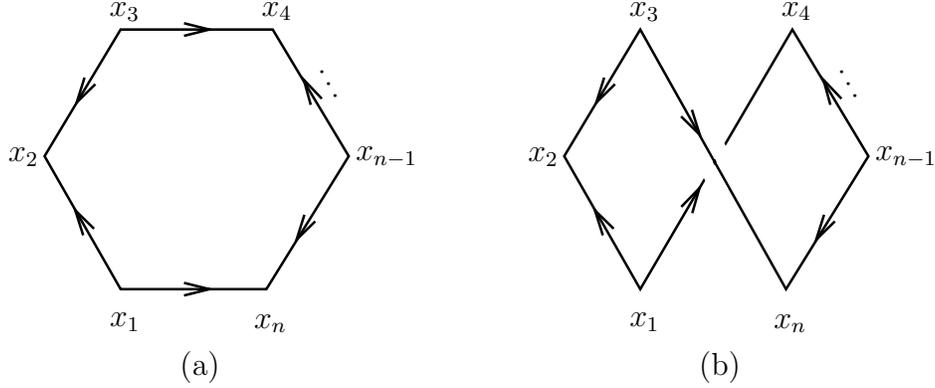} }
 \caption{\small  Feynman diagrams of different types contributing to the correlator \re{defco'} at tree level. Arrowed lines denote free scalar propagators $\vev{\bar \phi^{12}(x_i) \phi_{12}(x_j)}$.
 In the light-cone limit $x_{i,i+1}^2\to 0$ the leading
 contribution comes from diagram (a), while that of diagram
 (b) is suppressed by the factor $x_{34}^2 x_{1n}^2/(x_{3n}^2x_{14}^2)$.
 \label{ttree}}
 \end{figure}
This   can be remedied by considering the ratio
\begin{equation}\label{ratonc'}
    \lim_{x^2_{i,i+1}\to 0}G_n/G^{(0)}_n\,.
\end{equation}
Notice that the limit  \re{lim'} breaks the symmetry of $G_n$ under the  exchange of identical operators (e.g., exchanging all points with odd or with even numbers in \p{defco'}). Instead, it has a cyclic symmetry, $x_i \ \to\ x_{i+1}$, and a flip symmetry, $x_i \ \to\ x_{n-i+1}$.

{Secondly, the loop integrals develop additional light-cone singularities in the limit \re{lim'}.} To illustrate this, let us take a simple example -- the four-point correlator \p{defco'} at one loop. It has been computed in \cite{Eden:1998hh,GonzalezRey:1998tk,Eden:1999kh} and the result for the ratio $G_4/G^{(0)}_4$ is given by
\begin{equation}\label{exarat'}
    G_4/G_4^{(0)}  = 1 + 2 a\, x^2_{13}x^2_{24} g(x_i) + O(a^2)\,.
\end{equation}
Here the one-loop integral $g(x_i)$ is defined by
\begin{equation}\label{defg'}
g(x_i)   =   \frac{i}{2 \pi^{2}}
\int\frac{d^{4} x_0}{x^2_{10}   x^2_{20}   x^2_{30}   x^2_{40}}\,  .
\end{equation}
As long as the outer points are kept in generic positions, $x^2_{i,i+1} \neq 0$, this integral is finite and conformally covariant in four dimensions. This allows us to write it down as a function of two conformal cross-ratios
\begin{equation}\label{defg4'}
g(x_i)   =    \frac{1}{x_{13}^2 x_{24}^2}
\Phi^{(1)}(u,v) \,   , \quad u    =
\frac{x^2_{14} x^2_{23}}{x^2_{13} x^2_{24}}\,    , \quad v
  =   \frac{x^2_{12} x^2_{34}}{x^2_{13} x^2_{24}}\,    ,
\end{equation}
where the two-variable function $\Phi^{(1)}$ can be found in  \cite{davussladder}.
{The cross-ratios vanish in the limit \p{lim'} and
this function develops a logarithmic singularity, $\Phi^{(1)}(u,v)\sim \ln v \ln u$ as $u, v\to 0$.
It originates  from the integration in \re{defg'}, when $x_0$ approaches one
of the four light-cone segments $[x_i,x_{i+1}]$.

Therefore, to define the integral \re{defg'} in the limit \p{lim'} we have to introduce
a regularization. The standard approach is to use dimensional regularization from the very beginning,} that is, to repeat the whole calculation that leads to \p{exarat'}, but in $D=4-2\ep$ dimensions (with $\ep>0$). This approach was adopted in \cite{AEKMS}, where it was shown that the limit \p{ratonc'} turns the correlator into a light-like Wilson loop.

Alternatively, we might use the four-dimensional result \p{exarat'} and declare that we simply regularize the integral \p{defg'} by choosing a measure in $D=4-2\ep$ dimensions (with $\ep<0$). Notice the change of sign of the regulator -- now it looks more like an infrared, rather than the natural ultraviolet regulator needed for such short-distance singularities. This unusual choice is motivated by the observation that the one-loop space-time integral \p{defg'} is the dual space version \cite{Drummond:2006rz} of the one-loop scalar box momentum integral appearing in the four-gluon MHV amplitude. The latter is given by \cite{Bern:2005iz}
\begin{align}\label{MHV4'}
A_4/A_4^{(0)}= 1+ a\, s t I_4^{(1)}(p_i) + O(a^2)\,,
\end{align}
where $A_4^{(0)}$ is the tree-level amplitude,  $s=(p_1+p_2)^2$, $t=(p_3+p_4)^2$ are the Mandelstam variables and
\begin{align}
I_4^{(1)}(p_i)&=  \frac{2i}{(2\pi)^{2-2\ep}}  \int \frac{d^D k}{k^2(k-p_1)^2(k-p_1-p_2)^2(k+p_4)^2} \label{I1'}
\end{align}
is the one-loop scalar box integral.
Switching from momenta to  dual coordinates,
$k=x_{1}-x_{0}$ and
$p_i=x_i-x_{i+1}$ (with $x_5\equiv x_1$)\,,
we identify the two integrals:
\begin{align}
I_4^{(1)}(p_i) =   g_\ep(x_i) \,,
\end{align}
where {the subscript in} $g_\ep$ indicates that we have changed the measure in \p{defg'} to $D=4-2\ep$ dimensions (with $\ep<0$). Notice that the light-cone limit \p{ratonc'} for the correlator implies that the momenta $p_i=x_{i,i+1}$ are light-like, $p^2_i=0$, as required for the amplitude.

Further, comparing \p{exarat'} and \p{MHV4'}, we observe a surprisingly simple relation between correlator and amplitude:
\begin{equation} 
    \lim_{x^2_{i,i+1}\to 0}G_4/G_4^{(0)}  = \lr{A_4/{A}_4^{(0)}}^2 + O(a^2)\,.
\end{equation}

This one-loop exercise suggests the following general recipe for obtaining MHV gluon amplitudes from correlators of protected operators:
\begin{enumerate}
  \item Compute the $n-$point correlator of protected operators in four dimensions.
  \item Change the $D=4$ integration measure in the loop integrals to $D=4-2\ep$  (with $\ep<0$).
  \item Divide the correlator by its tree-level value and take the light-cone limit $x^2_{i,i+1} \to 0$.
  \item Switch from coordinates to dual momenta. The result should be the square of the $n-$gluon MHV amplitude, divided by the tree.
\end{enumerate}
Of course, the four-points/gluons case at one loop is probably too simple to allow us to jump to the conclusion that such a duality correlator/amplitude is a general phenomenon. It is the purpose of this paper to provide a lot of evidence in favor of this conjecture.

\section{Correlators of protected operators in superspace and their loop corrections}\label{Cpos}

In this section we explain the role of the Lagrangian insertion procedure in the calculation of loop corrections to correlators. Then we formulate the rules of dual infrared dimensional regularization and state the main result of the paper.

\subsection{Correlators of protected operators in $\cN=2$ harmonic superspace}

The one-loop, and even more so the two-loop  calculation of the correlators are much easier to do in terms of superspace Feynman graphs. Since we do not have an off-shell formulation of the $\cN=4$ SYM theory suitable for application to perturbation theory, the best compromise is to use the formulation in terms of $\cN=2$ superfields in the so-called harmonic superspace \cite{hh,Galperin:2001uw}. We give a brief summary of this formalism in Appendix~\ref{ApB}. Here we just mention the types of $\cN=2$ supermultiplets and superfields that we are dealing with.

The $\cN=4$ vector multiplet is decomposed into an $\cN=2$ matter multiplet (hypermultiplet) and an $\cN=2$ vector (gauge) multiplet. Upon reducing the R symmetry, $SU(4) \to SU(2)\times U(1)$, the six real scalars $\phi_{AB}$ split into an isodoublet $\phi^i$ (with $i=1,2$) and a complex singlet $\varphi$; the four (chiral) gluinos $\lambda^{A\a }$ split into a doublet $\lambda^{i\a }$ and two singlets $\psi^\a$, $\kappa^\a$ (and their antichiral conjugates). These fields can be combined to form $\cN=2$ superfields. One of them describes the hypermultiplet:
\begin{eqnarray}
q^+(x,\q^+, \bq^+,u) &=& \phi^i(x)u^+_i +\theta^{+\alpha}\psi_\alpha(x)
+\bar\theta^+_{\dot\alpha}\bar\kappa^{\dot\alpha}(x)+ \ldots
\label{onshq}
\end{eqnarray}
(the dots denote auxiliary and derivative terms).
It  is a Grassmann analytic (or G-analytic, or half-BPS) superfield in the sense that it depends on half of the Grassmann variables, $\theta^{+\alpha}=\q^{i\a}u^+_i$ and $\bar\theta^{+\dot\alpha}=\bq^{i\da}u^+_i$, obtained by projecting the $SU(2)$ doublets  $\q^{i\a}$ and $\bq^{i\da}$ with an $SU(2)$ harmonic variable $u^+_i$. The latter, together with its conjugate $u^-_i = (u^{+i})^*$ forms an $SU(2)$ matrix. Notice the presence of both chiral  and antichiral odd variables in the hypermultiplet superfield \p{onshq}. In harmonic superspace one can define a special conjugation (denoted by {tilde~} $\widetilde{}\,\, $), which takes \p{onshq} to another  G-analytic superfield,
\begin{eqnarray}
\tilde q^+(x,\q^+, \bq^+,u) &=& \bar\phi^i(x)u^+_i +\theta^{+\alpha}\kappa_\alpha(x)
+\bar\theta^+_{\dot\alpha}\bar\psi^{\dot\alpha}(x)+ \ldots\ .
\end{eqnarray}
In contrast with the hypermultiplet, the $\cN=2$ vector multiplet is described by the chiral field strength (and its antichiral conjugate)
\begin{equation}\label{chW}
    W(x,\q) = \varphi(x) + \q^{i\a}\lambda_{i\a}(x) + \q^{i\a} \q^\b_i F_{\a\b}(x) + \ldots
\end{equation}
containing, in particular, the self-dual part of the gluon field strength $F_{\a\b}=(\sigma^{\mu\nu})_{\a\b}  F_{\mu\nu}$.

The protected half-BPS bilinear operators \p{bil} in  the $SU(4)$ irrep $\mathbf{20'}$ split into a number of irreps of $SU(2)\times U(1)$. They can be descried as the lowest ($\q=0$) components of bilinears made of the  above superfields. Three such bilinears are G-analytic superfields:
\begin{align}\label{defO}
O= {\rm Tr}(q^+ q^+)\,,\qquad   \tilde O = {\rm Tr}(\tilde q^+
\tilde q^+)\,,\qquad  \hat O = \tilde{\hat O} = 2\, {\rm
Tr}(\tilde q^+  q^+)\,,
\end{align}
{where all operators are functions of $x,\q^+,\bq^+,u$}.
For example, the operator $O(x,\q^+,\bq^+,u)$ has the bottom
component $\cO(x,u) = O\vert_{\q^+=\bar\q^+=0} = {\rm
Tr}(\phi^i(x) \phi^j(x)) u^+_i u^+_j$ containing the complex
$SU(2)$ triplet ${\rm Tr}(\phi^i \phi^j)$. Another, real triplet
is the bottom component of the real operator $\hat O =\tilde{\hat
O}$, $\hat{{\cal O}}(x,u) = \hat{O}\vert_{\q^+=\bar\q^+=0} = 2 \,
{\rm Tr}(\bar\phi^i(x) \phi^j(x)) u^+_i u^+_j$. We may say that
the harmonic variables serve as a ``bookkeeping device" for
organizing the fields into $SU(2)$ representations.

The remaining  $SU(2)\times U(1)$ projections of the  $\mathbf{20'}$ are described by different types of superfields. Among them, the chiral operator ${\rm Tr}(WW)$ plays the prominent role of the $\cN=2$ SYM Lagrangian. In this paper we do not consider the rest of the $\cN=2$ projections of the  $\mathbf{20'}$: ${\rm Tr}(W \bar W)$, ${\rm Tr}(W q^+)$,  ${\rm Tr}(W \tilde q^+)$ and conjugates.

The correlators of the  three G-analytic operators \p{defO} will be the main subject of this and the following sections.
We are considering $n$-point correlators of the type, e.g.,
\begin{equation}\label{npt}
    G_n = \vev{{\cal O}(x_1,u_1)\tilde{\cO}(x_2,u_2) \cdots \cO(x_{n-1},u_{n-1}) \tilde{\cal O}(x_n,u_n)}
\end{equation}
for $n$ even, or
\begin{equation}\label{npt'}
    G_n = \vev{{\cal O}(x_1,u_1)\tilde{\cO}(x_2,u_2) \cdots \cO(x_{n-2},u_{n-2}) \tilde{\cal O}(x_{n-1},u_{n-1})\hat{\cO}(x_n,u_n)}
\end{equation}
for $n$ odd. At tree level the connected contribution to the
correlator has the form\footnote{More precisely, an $n$-point
correlator may involve $m$ complex operators ${\cal O}$ and also $m$ conjugate
operators $\tilde {\cal O}$, the remaining $n-2m$ operators being of the real type $\hat\cO$. At tree level, such a correlator  equals $c_{nm} \, G_n^{(0)}$, where
$c_{n0} = 1 + (-1)^n$ and $c_{nm} \, = \, (-1)^{n+m} \; : \, m > 0
$. In this article we are interested in the ratio of loop
corrections over the corresponding tree, which is universal. The
coefficients $c_{nm}$ can thus safely be omitted.}
\begin{equation}\label{Bornn}
    G^{(0)}_n = \frac{N_c^2}{(2\pi)^{2n}}\frac{(12)(23)\cdots(n1)}{x_{12}^2x_{23}^2 \cdots x_{n1}^2} + \mbox{subleading terms} \ ,
\end{equation}
where {$(r,r+1)$  is a shorthand for the $SU(2)$ invariant but $U(1)$ covariant contraction of the two
harmonics with labels $r$ and $r+1$,}
\begin{equation}\label{def(12)'}
  (r,r+1) = -(r+1,r) = u^{+i}_r \epsilon_{ij} u^{+j}_{r+1}\,.
\end{equation}
In \p{Bornn}, as in the expression  \p{tre''} for the tree amplitude for the highest-weight and lowest-weight state projections of the $\mathbf{20'}$, we only show the leading singular term in the light-cone limit \p{lim'}.  {Compared to \p{tre''}, the tree-level expression \p{Bornn} contains additional information about the isotopic $SU(2)$ structure of the correlator.} It is carried by the harmonic variables $u^{+\, i}_{r}$ (with $i=1,2$ and $r=1,\ldots,n$) at each point.

\subsection{Lagrangian insertions}

The correlators \p{npt} and \p{npt'} are defined by the path integral
\begin{align}\notag
    G_n  &= \int {\cal D}\Phi \ e^{iS_{\cN=4\ \rm SYM}}\  \cO(x_1,u_1)\ldots {\cal O}(x_n,u_n) \\
    &= G^{(0)}_n + g^2 G^{(1)}_n + g^4 G^{(2)}_n + \ldots\ . \label{patco}
\end{align}
Here the $\cN=4$ SYM action consists of two parts, the $\cN=2$ SYM action and the action of the $\cN=2$ hypermultiplet matter coupled to the gauge sector:
\begin{equation}\label{n4ac}
    S_{\cN=4\ \rm SYM} = S_{\cN=2\ \rm SYM} + S_{\cN=2\ {\rm matter}}\,.
\end{equation}

Instead of computing the loop corrections to $G_n$ directly, we prefer to evaluate the derivative with respect to the coupling. As we show below, it is given by the insertions of the $\cN=2$ SYM action
\begin{equation}\label{N2L}
    S_{\cN=2\ \rm SYM} = \int d^Dx d^4\theta\, L_{\cN=2\ \rm SYM}(x, \q)  = \int d^Dx d^4\bq\, \bar L_{\cN=2\ \rm SYM}(x, \bq)\,,
\end{equation}
where, after the appropriate rescaling of all the fields of the $\cN=2$  vector multiplet,
\begin{eqnarray}
  L_{\cN=2\ \rm SYM} &=& \frac{1}{2 g^2}{\rm Tr} (W^2) \label{chiL}\\
  &=& \frac{1}{2 g^2}\{ {\rm Tr}(\varphi^2)+ \ldots - (\q)^4{\rm Tr}[F_{\a\b} F^{\a\b} + 4
  \phi^i \square \bar\phi_i + 4 i\lambda^i \hat \pa \bar\lambda_i + \mbox{interaction terms} ]\}\nn
\end{eqnarray}
is the  $\cN=2$ SYM {\it chiral} Lagrangian. The effect of the rescaling is that the coupling $g$ disappears inside the Lagrangian $L_{\cN=2\ \rm SYM}$, it is only present in front of it, as indicated in \p{chiL}. The coupling also drops out from the interaction of the vector and matter multiplets (see Appendix~\ref{ip4p}).

Notice that the action in \p{N2L} has two forms, one chiral, the other antichiral. They are equivalent due to a Bianchi identity stating that the difference between ${\rm Tr} (W^2)$ and ${\rm Tr} (\bar W^2)$ is a {\it total (super)space derivative}. At the component level, this is clearly seen from \p{chiL}, where, for instance, the complex combination $F_{\a\b} F^{\a\b} = F^2 + i F \tilde F$ contains the  Yang-Mills Lagrangian $F^2$ and the topological term $i F \tilde F$.   We will come back to this important point later on.

Now, we want to differentiate the correlator with respect to the coupling, which is present only in $S_{\cN=2\ \rm SYM}$ as an overall factor, see \p{chiL}.
Thus, the one-loop (order $g^2$) correction to the correlator,
\begin{equation}
   g^2 \frac{\pa}{\pa g^2} \, G_n = - i \int d^D x_0 \, {\cal G}^{(0)}_{n+1}
   (x_0; x_1,u_1 ; \ldots ; x_n,u_n) + O(g^4) \label{g2}
\end{equation}
is calculated from the {\it Born-level}  $(n+1)$-point correlator
\begin{equation}\label{g41}
    {\cal G}^{(0)}_{n+1}(x_0; x_1,u_1 ; \ldots ; x_n,u_n) = \int d^4\q_0 \vev{ L_{\cN=2\ \rm SYM}(x_0, \q_0) \cO(x_1,u_1)\ldots {\cal O}(x_n,u_n)}  + O(g^4)\,,
\end{equation}
obtained by integrating the Lagrangian insertion over the Grassmann variables at the insertion point, but {not} over the space-time point $x_0$. Note that this tree-level correlator is of order $O(g^2)$, because it involves interaction vertices (see Sect.~\ref{n1l} for details).

{A very important property of the correlator ${\cal G}^{(0)}_{n+1}$ is its superconformal symmetry. Indeed, it involves the protected operators $\cO$ and $L_{\cN=2\  \rm SYM}$ (the latter belongs to the $\cN=4$ stress-tensor multiplet), with fixed conformal dimensions $2$ and $4$, respectively. Such operators are not renormalized and have well-defined conformal properties. This symmetry greatly facilitates the perturbative calculation, as explained in Appendix~\ref{ApB}.  }

The same procedure can be applied to the higher-order perturbative
corrections. Thus, to obtain the correlator at two loops (order
$g^4$), we compute the derivative
\begin{equation}
    \frac{1}{2} \, g^4 \left(\frac{\pa}{\pa g^2}\right)^2 G_n = - \frac{1}{2}
    \int d^D x_0 d^D x_{0'}\, {\cal G}^{(0)}_{n+2}(x_0, x_{0'}; x_1,u_1 ; \ldots ; x_n,u_n)   + O(g^6)  \label{g20}
\end{equation}
in terms of the Born-level  $(n+2)$-point correlator with two
Lagrangian insertions (see Appendix~\ref{ip4p})
\begin{equation}\label{g410}
    {\cal G}^{(0)}_{n+2}(x_0, x_{0'}; x_1,u_1 ; \ldots ; x_n,u_n) = \int d^4\q_0 d^4\q_{0'}\vev{ L(x_0, \q_0) L(x_{0'}, \q_{0'})  \cO(x_1,u_1)\ldots {\cal O}(x_n,u_n)} + O(g^6) \,.
\end{equation}

{In conclusion, the Lagrangian insertion procedure reduces the calculation of loop corrections to the correlator $G_n$ to a tree-level calculation of the correlator with insertions. The main point of our conjecture is that this tree-level correlator tells us what the {\it integrand} of the dual MHV amplitude should look like. The precise matching of the two objects is obtain by introducing a dual infrared regulator. }

\subsection{Outline of the dual infrared regularization procedure and of the duality correlator/amplitude}\label{fcair}

Our strategy for establishing the relationship between correlators and amplitudes is as follows.

We start with the tree-level correlators ${\cal G}^{(0)}_{n+1}$ \p{g41} and ${\cal G}^{(0)}_{n+2}$ \p{g410}. They are computed in $D=4$ and need no regularization. Then, we put ${\cal G}^{(0)}_{n+1}$ and ${\cal G}^{(0)}_{n+2}$ on the light cone by setting the adjacent external points at light-like distances, $x^2_{i, i+1} \to 0$, $i=1,\ldots,n$. In order to remove the pole singularities, we divide the correlator with insertions  by the tree-level correlator without insertions \p{Bornn}, thus obtaining $U(1)$ chargeless ratios, ${\cal G}^{(0)}_{n+1}/G_n^{(0)}$ and ${\cal G}^{(0)}_{n+2}/G_n^{(0)}$. After that it becomes safe to set the external points on the light cone, while keeping the insertion points $x_0$ and $x_{0'}$ in arbitrary positions. We remark that at this stage we still need no regularization.

Next, we perform the integration over the insertion points, thus passing from the tree-level correlators with insertions ${\cal G}^{(0)}_{n+1}$ and ${\cal G}^{(0)}_{n+2}$  to the loop corrections of the correlator without insertions $G_n$. Here we are facing logarithmic singularities due to the divergent integrals over $x_0$ and $x_{0'}$. Divergences arise when  the integration points approache {a light-like segment $[x_i, x_{i+1}]$.} We regularize the integrals by modifying the dimension of the integration measure, $D=4-2\ep$, with $\ep<0$. We emphasize that this is {\it not standard dimensional regularization}, for two reasons: (i) the tree-level correlator with extra points from the Lagrangian insertions has been computed in $D=4$ and then put on the light cone, without regularization; (ii) the sign of the regulator $\ep$ is chosen to match the {\it infrared divergences of the dual amplitude}, so this is not the usual ultraviolet regulator. We call this ``dual infrared regularization".

In order to make contact between the $n$-point correlator and the $n$-gluon amplitude, we identify the momenta with the dual coordinates
\begin{align}\label{dc1}
p_i=x_{i,i+1}\,,\qquad x_{n+1} \equiv x_1\,,
\end{align}
so that $\sum_1^n p_i=0$ and $p^2_i=0$. The Mandelstam variables are identified with the non-vanishing distances in dual space,
\begin{align}\label{dc2}
s_{ij}=(p_i+p_{i+1}+ \ldots + p_j)^2=x_{i,j+1}^2\,.
\end{align}

The main result of the next three sections is that the above procedure leads to the following duality relation between correlators restricted to the light cone and MHV $n$-gluon scattering amplitudes:
\begin{equation}\label{rellog}
    \lim_{x^2_{i,i+1}\to 0}\ \ln\lr{G_n/G_n^{(0)}} = \ln \lr{{A}_n/{A}_n^{(0)}}^2 + O(\ep)\,,
\end{equation}
where ${A}_n^{(0)}$ is the tree-level amplitude {and $O(\ep)$ denotes terms that vanish after we remove the regularization}.
The reason why we formulate the relation in terms of logs will become clear in Sect.~\ref{52loop}. In Sect.~\ref{n1l} we demonstrate this duality for any $n$ at one loop, and in Sections \ref{fpcfga} and \ref{52loop} we will show it for $n=4,5$ up to two loops.

\section{From correlators to amplitudes: $n$ points at one loop}\label{n1l}

In this section we perform the calculation of the one-loop correction to the correlator $G_n$ according to the procedure of Sect.~\ref{fcair}. \footnote{Our result agrees with earlier calculations of one-loop correlators in \cite{Eden:1998hh,GonzalezRey:1998tk,Eden:1999kh,Drukker:2009sf}.} We use the Feynman rules from Appendix~\ref{FR}. The harmonic superspace Feynman diagrams for the correlator with one extra point corresponding to the Lagrangian insertion are shown in Fig.~\ref{2loopgraphs}. \footnote{Notice that the graphs are drawn with a polygonal matter frame. Graphs based on the ``zigzag" configurations like in Fig.~\ref{ttree}(b) are suppressed in the light-cone limit \p{lim'}, after dividing out the leading singularity of the tree-level correlator \p{tre''}. \label{zigf}}

\begin{figure}[h!]
\psfrag{3}[cc][cc]{$k$}
\psfrag{4}[cc][cc]{$k+1$}
\psfrag{2}[cc][cc]{ }
\psfrag{1}[cc][cc]{$l+1$}
\psfrag{n}[cc][cc]{$ l$}
\psfrag{n-1}[cc][cc]{ }
\psfrag{0}[cc][cc]{$0$}
\psfrag{dots}[cc][cc]{ }
 \psfrag{(a)}[cc][cc]{(a)}
 \psfrag{(b)}[cc][cc]{(b)}
%
\centerline{ \includegraphics[height=60mm]{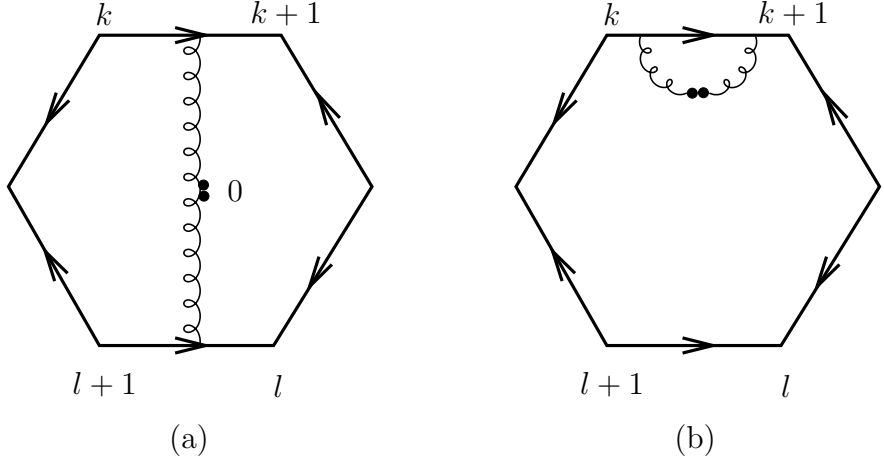} }
 \caption{\small  The $n$-point correlator with one insertion. The solid and wavy lines are hypermultiplet and gauge propagators, respectively.  The double dot denotes the insertion of the $\cN=2$ SYM  Lagrangian ${\rm Tr}(W^2)$. \label{2loopgraphs}  }
 \end{figure}

\noindent They  are constructed from the two basic building blocks
T and TT described in Appendix~\ref{Fgha}, combined with free
hypermultiplet propagators \p{qproat0}. The block T shown in
Fig.~\ref{blockI}(a) is the supersymmetric analog of the vertex
correction $\vev{\bar\phi_a(x_1) F_b^{\mu\nu}(x_0) \phi_c(x_{2})}$ for two scalars and one gauge field strength. The main difference is that at
the insertion point we have only the self-dual part of the
gauge field strength $F_{\a\b} = (\sigma^{\mu\nu})_{\a\b}
F_{\mu\nu}$, as well as the auxiliary field $Y^{ij}$ (see \p{6.10.1}); the scalar $\varphi$ and the gluino $\lambda^{i\a}$ from \p{chW} do not have a cubic vertex with
the external scalars. Since  we are doing the calculation in
$D=4$, the integral at the vertex can be easily computed
yielding a rational expression. It is obtained from the block
\p{T} by setting the external $\q^+_1=\q^+_2=0$:
\begin{align}
\vev{\tilde q^+_a(x_1,0,u_1) W_b(x_0, \q_0) q^+_c(x_2,0,u_2) } &=
- \frac{2 i g^2 f_{abc} }{(2\pi)^4}\frac{(12)}{x_{12}^2} i_{12}
\end{align}
with
\begin{align}\label{iblock}
i_{12}=x_{12}^2 \frac{\theta^+_{0/1}\cdot \theta^+_{0/2}}{(12)x_{10}^2 x_{20}^2}-\frac{\theta^+_{0/1}\cdot \theta^-_{0/1}}{ x_{10}^2}
+\frac{\theta^+_{0/2}\cdot \theta^-_{0/2}}{ x_{20}^2}-
\frac{\theta_{0/1}^+ [x_{10},x_{20} ] \theta_{0/2}^+}{(12) x_{10}^2x_{20}^2}\,.
\end{align}
Here $\theta^\pm_{0/r} = \theta_0^i (u_r)^\pm_i$ are the two $U(1)$ projections of the $SU(2)$ doublet $\theta_0^i$ with the harmonics at point $r$; $\q \cdot  \q \equiv \q^\a \q_a$; $[x,y ]_\a{}^\b \equiv x_{\a\da} y^{\da\b} - y_{\a\da} x^{\da\b} = -2i x^\mu y^\nu (\sigma_{\mu\nu})_\a^\b\,$.

 \begin{figure}[h]
 \psfrag{0}[cc][cc]{$0$}
 \psfrag{1}[cc][cc]{$1$}
 \psfrag{2}[cc][cc]{$2$}
 \psfrag{(a)}[cc][cc]{(a)}
 \psfrag{(b)}[cc][cc]{(b)}
\centerline{ \includegraphics[height=40mm]{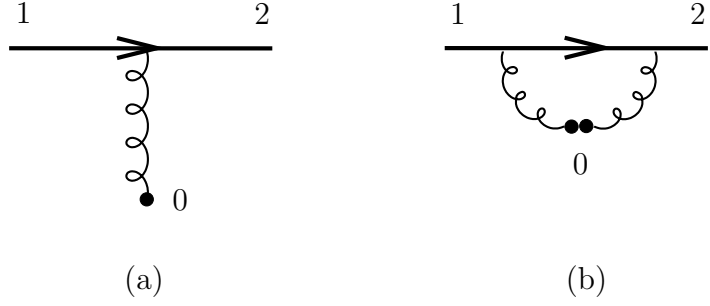} }
 \caption{\small Building blocks for the graphs in Fig.~\ref{2loopgraphs}. The single dot in (a) denotes a $W$ insertion, the double dot in (b) a ${\rm Tr}(W^2)$ insertion. \label{blockI}  }
 \end{figure}

The second building block shown in Fig.~\ref{blockI}(b)  {is the supersymmetric analog of the propagator correction $\vev{\bar\phi(x_k) \phi(x_{k+1})}$, after integration over the insertion point.} It is obtained from the TT block \p{TT} by identifying the insertion points $0\equiv 0'$ and by setting $\q^+_1=\q^+_2=0$:
\begin{equation}\label{defj12}
   \vev{\tilde q^+_a(x_1,0,u_1) {\rm Tr}(W^2)(x_0, \q_0) q^+_b(x_2,0,u_2) } =
    {2 g^4 N_c \, \delta_{ab}\over (2\pi)^6}{(12)\over x_{12}^2}j_{12}
\end{equation}
with
\begin{align}\label{defj12'}
j_{12} \equiv - {(1^-2^-)\over (12)}\frac{(\theta^+_{0/1})^2 (\theta^+_{0/2})^2}{x_{10}^2 x_{20}^2}= \half ( i_{12} )^2 \,.
\end{align}

Summing up all graphs (all sums are cyclic, $k+n \equiv k$) and integrating over the odd variable  $\q_0$ at the insertion point (but not yet over the even $x_0$), we find
\begin{eqnarray}
  {\cal G}_{n+1} &=&
{2 \, g^2 N^3_c\over (2\pi)^{2(n+2)}} {(12)(23)\cdots(n1)\over
x_{12}^2x_{23}^2 \cdots x_{n1}^2} \int d^4\theta_0\ \Bigl(
\frac{1}{2} \sum_{k, l=1;\, k \neq l}^n i_{k,k+1} i_{l,l+1} +
\sum_{k=1}^n j_{k,k+1} \Bigr) \nn\\ &=& \frac{a}{2 \pi^2} \,
G^{(0)}_n\ \int d^4\theta_0\ \Bigl(\sum_{k=1}^n
i_{k,k+1}\Bigl)^2\,, \label{sum2loop}
\end{eqnarray}
where $G^{(0)}_n$ is the leading singular part of the connected tree-level $n$-point correlator defined in \p{Bornn}. In deriving \p{sum2loop} we have used the identity $j_{kl} = \half (i_{kl}) ^2$.

The next step is to set the adjacent external points on the light cone, $x^2_{k,k+1} =0$, for the ratio
\begin{equation}\label{onlco}
    \lim_{x^2_{i,i+1}\to 0} {{\cal G}_{n+1}}/{G^{(0)}_n}  = \frac{a}{2 \pi^2}
 \lim_{x^2_{i,i+1}\to 0}\int   d^4\theta_0\ \Bigl(\sum_{k=1}^n i_{k,k+1}\Bigl)^2 \,.
\end{equation}
We notice that for $x_{12}^2\to 0$ the first term in $i_{12}$ \p{iblock} vanishes while the second and the third terms cancel in the cyclic sum
\begin{align}
\sum_{k=1}^n i_{k,k+1} = -\sum_{k=1}^n \frac{\theta_{0/k}^+ [x_{k,0},x_{k+1,0} ] \theta_{0/k+1}^+}{(k,k+1) x_{k,0}^2x_{k+1,0}^2}\,.
\end{align}
Then, using the identity
\begin{align}\label{id0}
  \int d^4\theta_0\, \theta_0^{\alpha_1 i_1}\theta_0^{\alpha_2 i_2}\theta_0^{\alpha_3 i_3}\theta_0^{\alpha_4 i_4}
= - \frac{1}{4} \left(
\epsilon^{i_1i_2}\epsilon^{i_3i_4}\epsilon^{\alpha_1\alpha_4}
\epsilon^{\alpha_2\alpha_3}
-
\epsilon^{i_1i_4}\epsilon^{i_2i_3}\epsilon^{\alpha_1\alpha_2}
\epsilon^{\alpha_3\alpha_4} \right)\,,
\end{align}
after some algebra we obtain
\begin{align} \notag
\lim_{x^2_{i,i+1}\to 0}& \int d^4\theta_0 \,  \lr{\sum_{k} i_{k,k+1}}^2
\\
& = \lim_{x^2_{i,i+1}\to 0}\sum_{k,l} \frac{[(k,k+1)(l,l+1)]^{-1}}{
 x_{k,0}^2x_{k+1,0}^2 x_{l,0}^2x_{l+1,0}^2}
 \int d^4\theta_0 \, \theta_{0/k}^+ [x_{k,0},x_{k+1,0} ] \theta_{0/k+1}^+\theta_{0/l}^+ [x_{l,0},x_{l+1,0} ] \theta_{0/l+1}^+  \notag\\
 & = \lim_{x^2_{i,i+1}\to 0}\Big\{ 2  \sum_{k,l} \frac{ (x_{k,0} \cdot x_{l+1,0})(x_{k+1,0}
 \cdot x_{l,0})-(x_{k,0}\cdot x_{l,0})(x_{k+1,0}\cdot x_{l+1,0})}
 {x_{k,0}^2x_{k+1,0}^2 x_{l,0}^2x_{l+1,0}^2}  \notag\\
 & \hspace*{70mm} + i \sum_{k,l} \frac{\ep_{\mu\nu\lambda\rho} x_{k,0}^\mu x_{k+1,0}^\nu x_{l,0}^\lambda x_{l+1,0}^\rho}
 {x_{k,0}^2x_{k+1,0}^2 x_{l,0}^2x_{l+1,0}^2} \Big\} \,. \label{pseli}
\end{align}
The parity-odd terms in the last line deserve a special comment. They will subsequently be integrated over the insertion point $x_0$, according to \p{g2}. The resulting pseudo-scalar integral will depend on the four external points $x_{k}, x_{k+1}, x_{l}, x_{l+1}$, which is not sufficient to make a translation invariant pseudo-scalar. But in fact there is another reason why these terms are suppressed by the integral over the insertion point. By inspecting the component content of the inserted Lagrangian \p{chiL}, we see that the pseudo-scalar terms are due to the presence  of the topological term $iF\tilde F$. Such an insertion is a total space-time derivative with respect to the insertion point, so the integral must vanish. {The underlying reason why the loop corrections to the correlator $G_n$ cannot contain parity odd terms is that the fields $\phi$ in the operators $\cO$ can be treated as true scalars (see Appendix~\ref{pary}). } We will come back to this point in Sect.~\ref{52loop}.

Further, replacing $2(x_{k,0}\cdot x_{l+1,0})=x_{k,0}^2+x_{l+1,0}^2-x_{k,l+1}^2$, etc. in \p{pseli} and using the properties of the cyclic sum over $k$ and $l$, we obtain
\begin{align}
\lim_{x^2_{i,i+1}\to 0}\int d^4\theta_0 \, \lr{\sum_{k}
i_{k,k+1}}^2= \frac{1}{2} \lim_{x^2_{i,i+1}\to 0}\sum_{k,l} \frac{
x_{k,l+1}^2 x_{k+1,l}^2- x_{kl}^2 x_{k+1,l+1}^2}{
 x_{k,0}^2x_{k+1,0}^2 x_{l,0}^2x_{l+1,0}^2}\ .
\end{align}
Notice that the numerator vanishes when $k=l-1,\, l$ or $l+1$.

Up to now, we have done the entire {\it tree-level} calculation, including the light-cone limit, in $D=4$. The last step is to integrate over the insertion point, $\int d^Dx_0$, with $D=4-2\ep$ and $\ep <0$, in order to regularize the divergences occurring when the integration point approaches any of the light-like segments $[x_i, x_{i+1}]$. {We reiterate that this is not the natural UV regularization (for which $\ep>0$), but it is what we call a dual infrared regularization.} This yields
\begin{equation}\label{1loopnn}
  \lim_{x^2_{i,i+1}\to 0}   \left. \frac{\pa}{\pa a} \left( G_n/G^{(0)}_n \right) \right\vert_{a=0}
    = - \frac{ i}{4 \pi^2} \lim_{x^2_{i,i+1}\to 0} \sum_{k,l} \int d^D x_0  \frac{ x_{k,l+1}^2 x_{k+1,l}^2- x_{kl}^2 x_{k+1,l+1}^2}{
 x_{k,0}^2x_{k+1,0}^2 x_{l,0}^2x_{l+1,0}^2}\,.
\end{equation}
{To make a comparison with the amplitudes, it is instructive to rewrite
the one-loop integral in the right-hand side of \re{1loopnn} in terms of dual
momenta, Eqs.~\p{dc1} and \p{dc2}. Defining the $D-$dimensional loop momenta as $\ell=x_{0,k}$ we find that, for general $k$ and $l$ this {\it space-time} integral becomes the two-mass easy box {\it momentum} integral~\cite{Bern:1994zx}
\begin{align}
F(p,P,q,Q) = - \frac{i}{4 \pi^2} \int \frac{d^D \ell\, (P^2 Q^2 - (p+P)^2 (q+Q)^2)}{\ell^2 (\ell+p)^2 (\ell+p+P)^2(\ell-Q)^2}\,, \qquad (p^2=q^2=0)
\end{align}
evaluated for
$p=x_{k,k+1}$, $P= x_{k+1,l}$, $q = x_{l,l+1}$ and $Q = x_{l+1,k}$.
When rewritten in terms of these functions, the right-hand side of \re{1loopnn}
 coincides with twice the one-loop $n$-gluon MHV amplitude (see Eq.~(4.19) in \cite{Bern:1994zx})
 \begin{align}
 \lim_{x^2_{i,i+1}\to 0}   \left. \frac{\pa}{\pa a} \left( G_n/G^{(0)}_n \right) \right\vert_{a=0}
 = 2   A_n^{(1)}/A_n^{(0)}
    = 2  \sum F(p,P,q,Q) \,.
\end{align}}
In conclusion, the {\it correlator} calculated by means of a Lagrangian insertion and in dual infrared dimensional regularization, reproduces the {\it amplitude} in terms of momentum integrals.

\section{Four-point correlators and four-gluon amplitudes\\ to two loops}\label{fpcfga}

In this section we extend the one-loop duality from the previous section to two loops. We show that the logarithm of the correlator of four half-BPS operators of weight two (bilinears), when put on the light cone using the dual IR regularization procedure, becomes identical with the logarithm of the square of the four-gluon scattering amplitude.

\subsection{Four-point correlators of bilinear half-BPS operators}\label{fpcbhbo}

Let us consider the correlator of four protected $\cN=2$ half-BPS complex operators \p{defO}
\begin{equation}\label{44pt}
    G_4 = \vev{{\cal O}(x_1,u_1)\tilde{\cO}(x_2,u_2)\cO(x_3,u_3) \tilde{\cal O}(x_4,u_4)}\,.
\end{equation}
It has been computed up to two loops in \cite{Eden:2000mv}~\footnote{The terminology of  \cite{Eden:2000mv} is based on the topology of the Feynman graphs, rather than on the perturbative order of the correlator. Thus, the lowest order (Born level) contribution is called ``one-loop", and the contributions of order $g^{2(\ell-1)}$ are called ``$\ell$-loop".} (we summarize the computation in Appendix~\ref{ApB}):~\footnote{ The Born approximation \p{Born0} has a connected and a disconnected sectors (see   \cite{Eden:2000mv}), with different color factors. Here we show only the connected contribution to the tree-level correlator. Also, the color factors are given in the large $N_c$ approximation.  }
\begin{align}
G_4^{(0)} &= \frac{N_c^2}{(2\pi)^8}\frac{(12)(23)(34)(41)}{x_{12}^2x_{23}^2x_{34}^2x_{41}^2}\,,
\label{Born0}\\[2mm]
G_4^{(1)} &= \frac{N_c^2}{(2\pi)^8} \frac{2 a\,\mathcal{R}}{x_{12}^2x_{23}^2x_{34}^2x_{41}^2}\ x_{13}^2 x_{24}^2 g_0(1,2,3,4)\,,
\label{Born1}\\[2mm]
G_4^{(2)} &= \frac{N_c^2}{(2\pi)^8}\frac{2 a^2  \mathcal{R}}{x_{12}^2x_{23}^2x_{34}^2x_{41}^2}\bigg[  \frac12(x_{12}^2x_{34}^2+x_{13}^2x_{24}^2+x_{14}^2x_{23}^2)
(g_0(1,2,3,4))^2\notag
\\[2mm] \notag
& \qqqquad
+  x_{13}^2 x_{24}^2 \bigg( x_{12}^2 h_0(1,2,3;1,2,4) +  x_{23}^2 h_0(1,2,3;2,3,4) +  x_{34}^2 h_0(1,3,4;2,3,4)
\\ \label{Born2}
&\qqqquad +  x_{41}^2 h_0(1,2,4;1,3,4)
+ x_{13}^2 h_0(1,2,3;1,3,4) + x_{24}^2 h_0(1,2,4;2,3,4)\bigg) \bigg]\,,
\end{align}
where the one- and two-loop four-dimensional integrals in {\it coordinate} space are defined by
\begin{align}  \label{h1}
g_0(1,2,3,4) &=c_0 \int\frac{d^4 x_0}{x_{10}^2x_{20}^2x_{30}^2x_{40}^2}\,,
\\  \label{h12}
h_0(1,2,3;1,2,4) &= c_0^2 \int \frac{d^4 x_0 d^4 x_{0'}}{(x_{10}^2x_{20}^2x_{30}^2 )x_{0{0'}}^2( x_{1{0'}}^2x_{2{0'}}^2x_{4{0'}}^2)}\,,
\end{align}
with $c_0=i/(2\pi^2)$.
A characteristic feature of the all-order loop corrections is the presence of the universal harmonic-space-time polynomial prefactor
\begin{align}\label{R''}
\mathcal{R} = (12)^2 (34)^2 x_{14}^2 x_{23}^2+(14)^2(23)^2 x_{12}^2x_{34}^2+(12)(23)(34)(41)\left[x_{13}^2x_{24}^2-x_{12}^2x_{34}^2-x_{14}^2x_{23}^2 \right]\,.
\end{align}
This phenomenon was first revealed in \cite{Eden:2000bk} under the name ``partial non-renormalization".

Note that if the external points are in generic positions with $x_{ij}^2\neq 0$, the integrals above are well defined in  $D=4$ and are manifestly conformally covariant.  This allows us to write them as functions of the conformal cross-ratios,
\begin{align}\notag
g_0(1,2,3,4)  &=  \frac{1}{x_{13}^2x_{24}^2} \Phi^{(1)}\lr{\frac{x_{12}^2x_{34}^2}{x_{13}^2x_{24}^2},\frac{x_{14}^2x_{23}^2}{x_{13}^2x_{24}^2}}\,,
\\ \label{6.10}
h_0(1,2,3;1,2,4) &= {1\over
(x_{12}^2)^2x_{34}^2}\Phi^{(2)} \Bigl({x_{14}^2x_{23}^2\over
x_{12}^2x_{34}^2}, {x_{13}^2x_{24}^2\over x_{12}^2x_{34}^2}\Bigr)\,,
\end{align}
having well-known expressions in terms of polylogs (see, e.g., \cite{davussladder}). In what follows we wish to take the light-cone limit $x^2_{i,i+1} \to 0$, which makes the integrals diverge. We regularize them, mimicking  the gluon scattering amplitudes,  by modifying the dimension of the integration measure to $D=4-2\ep$ (with $\ep<0$), thus giving up conformal invariance
\begin{align}\notag
g(1,2,3,4) &= c_\ep  \int \frac{d^Dx_0}{x_{10}^2 \, x_{20}^2 \, x_{30}^2 \,
x_{40}^2}\,,
\\
h(1,2,3;1,2,4) &= c_\ep^2 \int \frac{d^Dx_0 \, d^Dx_{0'}}{(x_{10}^2 \, x_{20}^2 \,
x_{30}^2) \, x^2_{0{0'}} \, (x_{1{0'}}^2 \, x_{2{0'}}^2
x_{4{0'}}^2)} \,,  \label{inth}
\end{align}
where the normalization factor $c_\ep= 2i/(2\pi)^{2-2\ep}$ is introduced to simplify the final expressions for the two-loop corrections.

\subsection{Four-gluon amplitudes}

We expect that the four-point correlator \re{44pt} in the  light-cone limit \re{lim'} is related
to the four-gluon planar MHV amplitude. The latter has been computed to
two loops in \cite{Bern:2005iz} and it has the following form:
\begin{align}\label{MHV4}
A_4= A_4^{(0)}\left[ 1+ a  M^{(1)} + a^2   M^{(2)} + O(a^3)\right]\,,
\end{align}
where $A_4^{(0)}$ is the tree-level amplitude and the loop corrections $M^{(1)}$ and $M^{(2)}$ are given by
\begin{align} 
M^{(1)} =  st I_4^{(1)}(s,t)\,,\qquad
M^{(2)} =    st\lr{sI_4^{(2)}(s,t)+tI_4^{(2)}(t,s) }\,,
 \label{tlin'}
\end{align}
where $s=(p_1+p_2)^2$ and $t=(p_2+p_3)^2$ are the Mandelstam variables.
All gluons are outgoing, so that their momenta satisfy the
relations $\sum_1^4 p_i =0$ and $p_i^2=0$.

The one- and two-loop scalar box momentum integrals $I_4^{(1)}$ and $I_4^{(2)}$ in \p{tlin'} are given by
\begin{align}
I_4^{(1)}(s,t)&=  c_\ep  \int \frac{d^D k}{k^2(k-p_1)^2(k-p_1-p_2)^2(k+p_4)^2} \label{I1}\,,
\\[2mm]
I_4^{(2)}(s,t)&= c_\ep^2  \int \frac{d^D k\,d^D l}{k^2(k-p_1)^2(k-p_1-p_2)^2(k+l)^2 l^2(l-p_4)^2(l-p_3-p_4)^2}\,. \label{I2}
\end{align}
These massless integrals are infrared divergent, therefore they are regularized dimensionally, $D=4-2\ep$ with $\ep<0$.
Switching from momenta to  dual coordinates \cite{Drummond:2006rz},
$k=x_{1}-x_{0}$, $l=x_{0'}-x_{1}$ and
$p_i=x_i-x_{i+1}$ (with $x_5\equiv x_1$)\,,
we identify the pseudo-conformal integrals from the  correlator with those from the amplitude (see Fig.~\ref{1linte}):
\begin{align}
I_4^{(1)}(s,t) =   g(1,2,3,4) \,,\qquad  I_4^{(2)}(s,t)=    h(1,2,3;1,3,4)\,,\qquad   I_4^{(2)}(t,s) =   h(1,2,4;2,3,4)\,.
\end{align}
Then, the two-loop expression for the amplitude \re{MHV4} can be rewritten as
\begin{align}\label{A4-ratio}
A_4/A_{4}^{(0)} = 1+ a \, x_{13}^2 x_{24}^2 g(1,2,3,4) + a^2 \, x_{13}^2 x_{24}^2 \big[
x_{13}^2\, h(1,2,3;1,3,4) + x_{24}^2 \, h(1,2,4;2,3,4)\big]\,.
\end{align}

\begin{figure}[h!]
\psfrag{h1}[cc][cc]{$g(1,2,3,4)\ \leftrightarrow \ I_4^{(1)}(s,t)$}
\psfrag{h2}[cc][cc]{$h(1,2,3;1,3,4)\ \leftrightarrow \ I_4^{(2)}(s,t)$}
\psfrag{0}[cc][cc]{$\scriptstyle 0$}
\psfrag{0'}[cc][cc]{$\scriptstyle 0'$}
\psfrag{1}[cc][cc]{$\scriptstyle 1$}
\psfrag{2}[cc][cc]{$\scriptstyle 2$}
\psfrag{3}[cc][cc]{$\scriptstyle 3$}
\psfrag{4}[cc][cc]{$\scriptstyle 4$}
\psfrag{1i}[cc][cc]{$\scriptstyle \it 1$}
\psfrag{2i}[cc][cc]{$\scriptstyle \it 2$}
\psfrag{3i}[cc][cc]{$\scriptstyle \it 3$}
\psfrag{4i}[cc][cc]{$\scriptstyle \it 4$}
\centerline{ \includegraphics[width=0.7\textwidth]{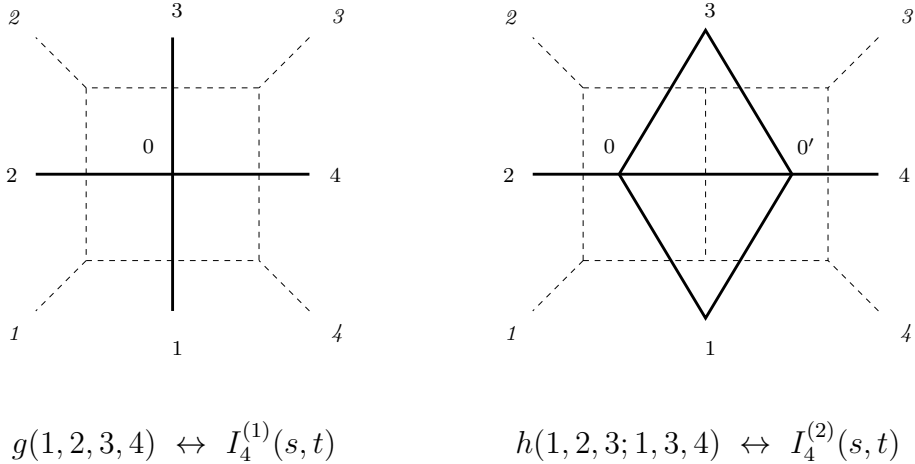} }
 \caption{\small  One- and two-loop pseudo-conformal integrals
 contributing to the correlator $G_4$, Eq.~\re{44pt}, and to the amplitude $A_4$, Eq.~\re{MHV4}. The diagrams with solid lines depict Feynman integrals in $x-$space.
The diagrams with dashed lines represent the same integral in the dual  momentum space. The straight labels correspond to the points $x_i$,  the slanted labels correspond to the momenta  $p_i=x_i-x_{i+1}$.  \label{1linte}}
 \end{figure}

\subsection{Duality correlator/amplitude}

For the purpose of comparing correlators and amplitudes, we define the ratio of the correlator and its Born-level expression:
\begin{align}  \label{rati}
G_4/G_4^{(0)} =  1 + &  \frac{\mathcal{R}}{(12)(23)(34)(41)}
\\
\times &
\bigg\{ 2a \, x_{13}^2 x_{24}^2 g(1,2,3,4) +
a^2\bigg[ (x_{12}^2x_{34}^2+x_{13}^2x_{24}^2+x_{14}^2x_{23}^2)
(g(1,2,3,4))^2\notag
\\[1mm] \notag
& 
+ 2x_{13}^2 x_{24}^2 \bigg(  x_{12}^2 h(1,2,3;1,2,4) +   x_{23}^2 h(1,2,3;2,3,4) +   x_{34}^2 h(1,3,4;2,3,4)
\\[1mm] \notag
&  +  x_{41}^2 h(1,2,4;1,3,4)
+ x_{13}^2 h(1,2,3;1,3,4) + x_{24}^2 h(1,2,4;2,3,4)\bigg)\bigg]+O(a^3)\bigg\}\,.
\end{align}
Next, we wish to evaluate this ratio  on the light cone, i.e. with
\begin{equation}\label{oncone}
    x^2_{12}=x^2_{23}=x^2_{34}=x^2_{41}=0\,.
\end{equation}
From \p{R''} we see that the prefactor ${\mathcal R}/[(12)(23)(34)(41)]$ in \p{rati} is reduced to $x^2_{13}x^2_{24}$. Further, most of the two-loop $h-$integrals  do not contribute to the right-hand side of \p{rati} due to vanishing kinematic prefactors like $x^2_{12}$. The result is
\begin{align}\notag
    \lim_{x^2_{i,i+1}\to 0} G_4/G_4^{(0)}  = 1 + & 2 a\, x^2_{13}x^2_{24} g(1,2,3,4)
    +a^2 \bigg[ \big(x^2_{13}x^2_{24} g(1,2,3,4)\big)^2
\\ & \label{ratio}
+ 2 x_{13}^2 x_{24}^2\bigg( x_{13}^2 h(1,2,3;1,3,4) + x_{24}^2 h(1,2,4;2,3,4)\bigg) \bigg]+O(a^3)\,.
\end{align}
Comparing this result with the perturbative expansion of the amplitude, Eq.~\re{A4-ratio}, we obtain 
\begin{equation}\label{relca4p}
    \lim_{x^2_{i,i+1}\to 0}G_4/G_4^{(0)}  = \lr{A_4/{A}_4^{(0)}}^2 + O(a^3)\,,
\end{equation}
confirming the general statement \p{rellog}.

{We recall that $G_4$ was defined in \re{44pt} as the correlation function
of operators \re{o2} bilinear in the scalar fields. The duality \re{relca4p} can be extended to correlators of half-BPS operators $\cO={\rm Tr}(\phi^k)$ of arbitrary weight $k$, see Appendix \ref{multiap}.

\section{Five-point correlators and five-gluon amplitudes\\ to two loops}\label{52loop}

In this section we extend the previously found duality between dual-IR-regularized correlators and gluon MHV amplitudes to the case of five points/gluons. This is a rather non-trivial test, in view of the significantly more complicated integrals involved. Also, we explain why the duality should be formulated in terms of logs, rather than the correlator/amplitude themselves. The reason is in the parity-odd (pseudo-scalar sector) of the amplitude, which is reduced to $O(\ep)$  terms by taking the log. This is essential for the duality to work, because the correlator must be a true scalar, as we argue in Appendix~\ref{pary}.

We consider the correlator of $n=5$  half-BPS operators of weight $k=2$,
\begin{align}\label{G5}
G_5 =\vev{\cO(1) \tilde\cO(2)\cO(3) \tilde\cO(4) \hat{\cO}(5) }\,,
\end{align}
where the local scalar operators $\cO$, $\tilde\cO$ and $\hat{\cO}$ are the bottom components of the hypermultiplet bilinears
\begin{align}
O= {\rm Tr}(q^+ q^+)\,,\qquad   \tilde O = {\rm Tr}(\tilde q^+
\tilde q^+)\,,\qquad  \hat{O} = 2 \, {\rm Tr}(\tilde q^+ q^+)\,,
\end{align}
and $i=(x_i,u_i^+)$ denotes the set of space-time and harmonic coordinates  of the hypermultiplet scalar fields,
$q^+(i)|_{\q=0} = \phi^r (x_i) u_{ir}^{+}$ and $\tilde q^+(i)|_{\q=0} = \bar \phi^r (x_i) u_{ir}^{+}$ (with $r=1,2$).
We wish to examine the correlator \re{G5} in
the limit
\begin{align}\label{lc}
x_{i,i+1}^2\equiv (x_i-x_{i+1})^2 \to 0\,,\qquad x_{i+5}\equiv x_i\,.
\end{align}
In this limit, the leading
asymptotic behavior of the tree-level correlator is given by the product of free scalar propagators
\begin{align}
G^{(0)}_5  {=}   \frac{N_c^2}{(2\pi)^{10}}\frac{(12)(23)(34)(45)(51)}{x_{12}^2 x_{23}^2 x_{34}^2 x_{45}^2 x_{51}^2}  + \ldots\,,
\end{align}
where the ellipses denote subleading terms as $x_{i,i+1}^2\to 0$. Notice that
the obvious symmetry of the correlator \re{G5} under the exchange of
operators, $1\leftrightarrow 3$ and $2\leftrightarrow 4$, is lost in the light-cone limit.
At loop level,
the correlator \re{G5} turns out to have the same leading singularity as the tree $G^{(0)}_5$. This suggests to study the following ratio in the light-cone limit \re{lc},
\begin{align}\label{F}
F_G\equiv  \lim_{x_{i,i+1}^2\to 0}\ln \left({G_5}/{G^{(0)}_5}\right)
=   a F_G^{(1)}(x_i) + a^2  F_G^{(2)}(x_i) + O(a^3)\,,
\end{align}
where $a=g^2N_c/(8\pi^2)$ and $F_G^{(p)}(x_i)$ are scalar functions of $x_i$ only.
The rationale for considering the log  of the ratio of correlators
in the left-hand side of \re{F} is that, firstly, it does not receive a contribution at $O(a^0)$ and, secondly, as we will argue below,  it has a much simpler form.
Computing $F_G^{(1)}(x_i)$ and $F_G^{(2)}(x_i)$ we shall follow the same routine as before, that is,
we shall expand  $F_G^{(p)}(x_i)$ over a basis of one- and two-loop pseudo-conformal integrals in $D=4$ dimensions, and then regularize them by
modifying the integration measure at the Lagrangian insertion points to $D=4-2\epsilon$ dimensions, $d^4 x_0\to d^{4-2\epsilon}x_0$, with $\ep<0$.

In Sect.~\ref{n1l} we have demonstrated  that the one-loop correction in \re{F} is given by
\begin{align}\label{fin1}
F_G^{(1)}(x_i)  =   x_{13}^2 x_{24}^2 g(1,2,3,4) + \text{(cyclic)}\,,
\end{align}
where $g(1,2,3,4)$ is the one-loop pseudo-conformal ``cross" integral in \re{penta};  (cyclic) means the four non-trivial cyclic permutations, $i\mapsto i+1$, of the  points $\{1,2,3,4,5\}$. The two-loop correction
$F^{(2)}_G$ is computed in Appendix~\ref{ip5p}, using the method of double Lagrangian insertions in harmonic superspace. The result is the following expression: \footnote{We are grateful to Fernando Alday for pointing out a misprint in \re{fin} in the original version of the paper.}
\begin{align}\notag 
F_G^{(2)}  = {}
 &   x^2_{13} \, x^2_{24} \, \left[
2 \, x^2_{13} \, h(1,2,3;1,3,4)+ 2 \, x^2_{24} \, h(1,2,4;2,3,4)-  x^2_{14} \, h(1,2,4;1,3,4) \right]
\\[2mm]
\notag
+ &  x^2_{13} \, x^2_{14} \, x^2_{25} \left[ 2 \, h(1,2,5;1,3,4) \, - \,
h(1,2,3;1,4,5) \, - \, h(1,2,4;1,3,5) \right]
\\[2mm] \notag
+ &  x^2_{24} \, x^2_{35} \left[ 2 \, x^2_{25} \, p(1;2,5;3,4) \, - \,
x^2_{24} \, p(1;2,4;3,5) \, - \, x^2_{35} \, p(1;3,5;2,4) \right]
\\[2mm] \label{fin}
-&  \ft14
 [x_{13}^2 x_{24}^2 g(1,2,3,4) ]^2 -  \ft12x^2_{13} \,x^2_{24} \, g(1,2,3,4) \  x^2_{13} \, x^2_{25} \, g(1,2,3,5)+   \mathrm{(cyclic)}\,,
\end{align}
where  the cyclic permutations act on the entire sum of terms. Here  we use the dimensionally regularized integrals \re{inth} together with
\begin{align}
\label{penta}
p(1;2,3;4,5) &= c_\ep^2 \int \frac{d^Dx_0 \, d^Dx_{0'}  \ x^2_{1{0'}}}{(x_{10}^2 \,
x_{20}^2 \, x_{30}^2) \, x^2_{0{0'}} \, (x_{2{0'}}^2 \, x_{3{0'}}^2 \, x_{4{0'}}^2 \,
x_{5{0'}}^2)}\,,
\end{align}
where $D=4-2\ep$ with $\ep<0$ and the normalization factor $c_\ep= 2i/(2\pi)^{2-2\ep}$ is introduced to simplify the final expressions for the two-loop corrections. Switching to dual momenta $p_i=x_i-x_{i+1}$, we find
that the $g-$, $h-$ and $p-$integrals correspond to the scalar box, double-box
and penta-box momentum integrals of Ref.~\cite{Bern:2006vw}, respectively.

\subsection{Five-gluon MHV amplitude}

We expect that in the light-cone limit \re{lc} the five-point correlator \re{G5} is related
to the five-gluon planar MHV amplitude. The latter has been computed to
two loops in \cite{Bern:2006vw} and it has the form
\begin{align}\label{MHV5}
A_5= A_5^{(0)}\left[ 1+ a  M^{(1)} + a^2   M^{(2)} + O(a^3)\right]\,,
\end{align}
where $A_5^{(0)}$ is the tree five-gluon MHV amplitude. Here the one- and two-loop corrections  are given
by the following expressions
\begin{align}\notag
M^{(1)}&=    \frac12 \sum_{\rm cyclic} s_{12} s_{23} I^{(1)}_a + M_1^{(\rm odd)}\,,
\\
M^{(2)}&=   \frac12\sum_{\rm cyclic} \left[s_{12}^2 s_{23} I^{(2)}_a+ s_{12}^2 s_{15} I^{(2)}_b + s_{12} s_{34} s_{45}  I^{(2)}_c \right] + M_2^{(\rm odd)}\,,
 \label{tlin}
\end{align}
where $s_{ij}=(p_i+p_j)^2$ is the invariant mass of the gluons with labels $i$ and $j$.
All gluons are considered outgoing, so that their momenta satisfy the
relations $\sum_1^5 p_i =0$ and $p_i^2=0$.

Further, $I^{(1)}_a$ is a one-loop
and $I^{(2)}_a$, $I^{(2)}_b$, $I^{(2)}_c$ are two-loop planar scalar integrals
depending on the gluon momenta, Eqs.~\re{I1} and \re{I2}. Switching to dual coordinates,
$p_i=x_i-x_{i+1}$, these integrals can be expressed in terms of the basis integrals
\re{penta}:
\begin{align}\notag
& I^{(1)}_a  =   g(1,2,3,4)
\,,
&&
 I^{(2)}_a  =    h(1,3,2;1,3,4)\,,
\\[3mm]
&  I^{(2)}_b =   h(1,3,2;1,3,5)
\,,
&&
 I^{(2)}_c   =   p(2;1,3;4,5)\,.
\end{align}

Finally,
$M^{(1)}_{\rm odd}$ and $M^{(2)}_{\rm odd}$ in \p{tlin} stand for the parity-odd contributions
to the five-gluon MHV amplitude, proportional to the pseudo-scalar $\epsilon(p_1,p_2,p_3,p_4)$.
The one-loop parity-odd contribution $M^{(1)}_{\rm odd}$ is of order $O(\ep)$, whereas  $M^{(2)}_{\rm odd}$
has a simple pole $1/\epsilon$. The residue at this pole is proportional
to the product of the one-loop parity-even and parity-odd parts. As a consequence,
the parity-odd contribution can be significantly simplified by considering the logarithm of the ratio $A_5/A_5^{(0)}$, as shown  in \cite{Bern:2006vw}. Namely, we introduce the following ratio function
\begin{align}\label{FA}
F_A\equiv  \ln \left(A_5/ A_5^{(0)}\right)^2 =
 a F_A^{(1)} +   a^2 F_A^{(2)} + O(a_\ep^3)\,, &
 \\[2mm] \notag
  F_A^{(1)}= 2M^{(1)}\,,\qquad F_A^{(2)} = 2 M^{(2)} - (M^{(1)})^2\,. &
\end{align}
The main advantage of \re{FA} is that, unlike the amplitude itself, the parity-odd contributions to $F_A^{(1)}$ and $F_A^{(2)}$ vanish as $\ep\to 0$.
Replacing $M^{(1)}$ and $M^{(2)}$ in \re{FA} by their explicit expressions \re{tlin} and
going to dual coordinates $p_i=x_i-x_{i+1}$, we obtain (up to terms vanishing
as $\epsilon \to 0$)
\begin{align} \label{amp1}
F_A^{(1)} &=   x_{13}^2 x_{24}^2 g(1,2,3,4) + \text{(cyclic)}
\\[3mm] \notag
F_A^{(2)}  &=   x_{13}^4 x_{24}^2 h(1,2,3;1,3,4)+
x_{24}^4 x_{13}^2 h(1,2,4;2,3,4) + x_{24}^2  x_{25}^2 x_{35}^2 p(1;2,5;3,4)
\\[2mm] \notag
&- \ft14 [x_{13}^2 x_{24}^2 g(1,2,3,4)]^2 - \ft12
x_{13}^2 x_{24}^2 g(1,2,3,4) x_{24}^2 x_{35}^2   g(2,3,4,5)
\\[2mm] \label{amp2}
&-\ft12 x_{13}^2 x_{24}^2 g(1,2,3,4) x_{35}^2 x_{14}^2   g(3,4,5,1)
+ \text{(cyclic)}\,.
\end{align}
Here $g$, $h$ and $p$ stand for the dimensionally regularized integrals defined in \re{penta}.

{Although the explicit expressions for individual two-loop integrals in \re{amp2} are quite complicated, their sum $F_A^{(2)}$ can be found in a closed
form thanks to the remarkable duality between MHV amplitudes and light-like Wilson loops \cite{am07,Drummond:2007aua,Brandhuber:2007yx}. This duality allows us to formulate the dual conformal Ward identity for $F_A$ to all loops \cite{Drummond:2007au}. For five points it  has a unique solution (up to an additive constant) which coincides with the BDS ansatz \cite{Bern:2005iz}. }

\begin{figure}[h!]
\psfrag{h1}[cc][cc]{$h(1,2,3;1,3,4)$}
\psfrag{h2}[cc][cc]{$h(1,2,4;2,3,4)$}
\psfrag{h3}[cc][cc]{$h(1,2,4;1,3,4)$}
\psfrag{h4}[cc][cc]{$h(1,2,5;1,3,4)$}
\psfrag{h5}[cc][cc]{$h(1,2,3;1,4,5)$}
\psfrag{h6}[cc][cc]{$h(1,2,4;1,3,5)$}
\psfrag{h7}[cc][cc]{$p(1;2,5;3,4)$}
\psfrag{h8}[cc][cc]{$p(1;2,4;3,5)$}
\psfrag{h9}[cc][cc]{$p(1;3,5;2,4)$}
\psfrag{1}[cc][cc]{$\scriptstyle 1$}
\psfrag{2}[cc][cc]{$\scriptstyle 2$}
\psfrag{3}[cc][cc]{$\scriptstyle 3$}
\psfrag{4}[cc][cc]{$\scriptstyle 4$}
\psfrag{5}[cc][cc]{$\scriptstyle 5$}
\psfrag{1i}[cc][cc]{$\scriptstyle \it 1$}
\psfrag{2i}[cc][cc]{$\scriptstyle \it 2$}
\psfrag{3i}[cc][cc]{$\scriptstyle \it 3$}
\psfrag{4i}[cc][cc]{$\scriptstyle \it 4$}
\psfrag{5i}[cc][cc]{$\scriptstyle \it 5$}
\psfrag{m1i}[cc][cc]{$\scriptstyle \it -1$}
\psfrag{m2i}[cc][cc]{$\scriptstyle \it -2$}
\psfrag{m3i}[cc][cc]{$\scriptstyle \it -3$}
\psfrag{m4i}[cc][cc]{$\scriptstyle \it -4$}
\centerline{ \includegraphics[height=170mm]{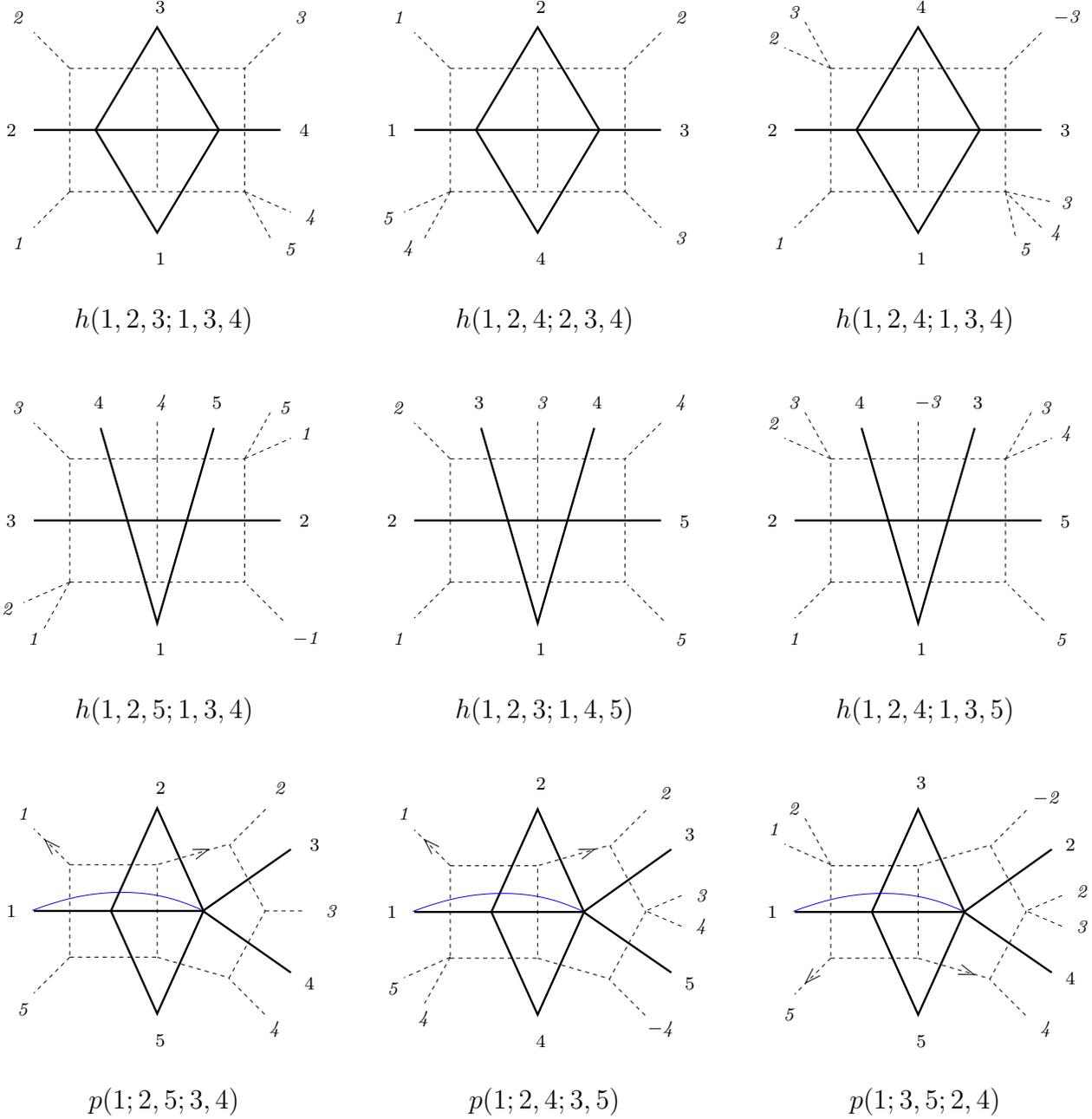} }
 \caption{\small  Two-loop pseudo-conformal integrals of different topologies
 contributing to the correlator $G_5$, Eq.~\re{fin}, and to the amplitude $A_5$, Eq.~\re{amp2}.
 The diagrams with solid lines depict Feynman integrals in $x-$space.
 The diagrams with dashed lines represent the same integral in the (dual)
 momentum space $p_i=x_i-x_{i+1}$. In the latter case,  $\scriptstyle (-k)$ stands for the particle with momentum $(-p_k)$. Thin solid lines denote numerators in the
 $x-$integral. In momentum space, this numerator  is given by the squared sum of the momenta flowing through the arrowed dashed lines.
 \label{fig0}}
 \end{figure}

\subsection{Duality correlator/amplitude}\label{Dure}

Assuming the duality relation \re{rellog}, the correlator on the light cone
should be related to the scattering amplitude
\begin{align}\label{dual1}
  \lim_{x_{i,i+1}^2\to 0}\ln \left(  {G_5}/{G^{(0)}_5}\right) = \ln \left( {A_5}/{A_5^{(0)}}\right)^2 + O(\ep) \,.
\end{align}
In terms of the functions $F_G$ and $F_A$ introduced in \re{F} and \re{FA}, this relation implies that up to terms vanishing for $\ep\to 0$
\begin{align}\label{dual2}
F_A^{(1)} = F_G^{(1)} \,,\qquad F_A^{(2)} = F_G^{(2)}\,,\quad \ldots
\end{align}
The first of these relations follows immediately from the explicit one-loop expressions \re{amp1} and \re{fin1}. At two loops, $F_A^{(2)}$ and $F_G^{(2)}$ are given by two
seemingly  different expressions, Eqs.~\re{amp2} and \re{fin}, respectively.
Then, for the duality relation \re{dual2} to hold, the two-loop integrals in \re{amp2} and \re{fin} should satisfy a very non-trivial identity.

To begin with, let us examine the topology of the Feynman integrals in the two-loop expression for $F_A^{(2)}$ and $F_G^{(2)}$. By construction, the planar MHV amplitude $A_5$ receives contributions from planar Feynman diagrams only. However, taking the log  of the two-loop amplitude \re{MHV5}, we find that
$F_A^{(2)}$ contains an admixture of products of one-loop integrals in \re{amp2}. For the correlator the situation is different. The diagrammatic
representation of $F_G^{(2)}$ is shown in Fig.~\ref{fig0}. We observe that the expression for $F_G^{(2)}$ involves two-loop integrals of the same type
as those contributing to $F_A^{(2)}$, Eq.~\re{amp2}. Closer examination shows, however, that
the external points in these integrals, $x_1,\ldots,x_5$, are not (anti)clock-wise  ordered and, therefore, these integrals do not contribute to the planar amplitude $F_A^{(2)}$. Still, they contribute to the two-loop planar correlator.
This is due to the different realizations of the planarity condition for correlators and scattering amplitudes. For the latter, the planar diagrams have the topology of a disk, while for the former they have the topology of a sphere (see Fig.~\ref{survivors} in Appendix~\ref{Fgha} for an example). Despite of this, the expected duality relation
$F_A^{(2)}=F_G^{(2)}$ suggests that the two sums of two-loop integrals of   different
topologies, Eqs.~\re{fin} and \re{amp2}, are equal to each other.

We can apply the relation $p_i=x_i-x_{i+1}$ to rewrite the Feynman integrals contributing to $F_G^{(2)}$ as conventional momentum integrals, as shown in Fig.~\ref{fig0}. The absence of
cyclic ordering of the external points $x_i$ leads to an unusual feature of
the resulting momentum integrals: They involve seven external legs including
one pair of legs with opposite momenta, $p_i$ and $(-p_i)$.

\subsubsection{Soft limit}

There exists a simple way to test the duality relations \re{dual1} and \re{dual2}
by examining them in the special limit $x_4\to x_5$. For the amplitude,
this corresponds, e.g., to the soft limit $p_4=x_{45}\to 0$ when one of the external
gluon momenta vanishes. In this limit, the dimensionally regularized five-gluon MHV amplitude $A_5$ reduces to the four-gluon amplitude $A_4$ in such a way that the ratio function \re{FA} takes the form
\begin{align}\label{A-soft}
\lim_{x_4\to x_5} F_A = \ln \lr{A_4 /A_4^{(0)}}^2 \,.
\end{align}
For the correlator \re{G5}, the limit $x_4\to x_5$ corresponds to its short
distance asymptotic behavior. In this limit, we apply the operator product expansion $\tilde\cO(4) \hat{\cO}(5)  \sim \vev{\bar \phi^+(4) \phi^+(5)} \tilde{\cO}(4) + \ldots$, where $\vev {\bar \phi \phi}$ is   a free scalar field propagator in $D=4-2\ep$ dimensions, and the dots denote subleading terms. In this way, we find from \re{F} that the ratio of the 5-point correlators reduces to the
ratio of the 4-point ones $G(1,2,3,4)=\vev{\cO(1) \tilde\cO(2)\cO(3) \tilde{\cO}(4) }$,
\begin{align}\label{G-soft}
\lim_{x_4\to x_5} F_G = \lim_{x_{i,i+1}^2\to 0}\ln \left( \frac{G(1,2,3,4)}{G^{(0)}(1,2,3,4)}\right) \,,
\end{align}
where the light-cone limit in the right-hand side corresponds to $x_{12}^2, x_{23}^2, x_{34}^2, x_{41}^2\to 0$.

Let us verify \re{A-soft} using the expression for the two-loop ratio function \re{FA}.
We recall that, by definition, the adjacent points in dual space are light-like separated, $x_{i,i+1}^2=0$. In the soft limit $x_4\to x_5$ we find the additional relations $x_{35}^2 \to 0$ and $x_{41}^2\to 0$. Notice that the dimensionally regularized $g-$, $h-$ and $p-$integrals, Eq.~\re{penta}, remain finite as $x_4\to x_5$.
The expression for $F_A$ simplifies because some of these integrals are multiplied by $x_{35}^2$ and $x_{41}^2$, so they do not contribute in the soft limit. Thus, the pentabox integral $p$ drops out:
\begin{align}\label{FA-s}
\lim_{x_4\to x_5} F_A^{(2)}  &=  2 \,x_{13}^4 x_{24}^2 h(1,3,2;1,3,4)+ 2\, x_{24}^4 x_{13}^2 h(2,4,1;2,4,3)
 -  [x_{13}^2 x_{24}^2 g(1,2,3,4)]^2\,.
\end{align}
Using Eq.~\re{A4-ratio}, we verify that the expression in the right-hand side of
\re{FA-s} coincides with the two-loop correction to the ratio of four-gluon amplitudes
$ 2 \ln (A_4 /A_4^{(0)})$, in agreement with \re{A-soft}.

In a similar manner, we examine the ratio function $F_G^{(2)}$ and simplify
the expression \re{fin} in the limit $x_4\to x_5$. We find that, unlike the case of
$F_A^{(2)}$, the pentabox integrals have a non-vanishing contribution which can be
expressed in terms of $h-$integrals. For instance,
\begin{align}
- x^2_{24} \, x^2_{35}  x^2_{35} \, p(1;3,5;2,4) +\text{(cyclic)} & \stackrel{x_4\to x_5}{\to}  - b(1,2,3,4)-x_{13}^4 x_{24}^2 h(1,3,2;1,3,4)\,,
\end{align}
where the notation was introduced for the `bad' integral
$b(1,2,3,4)=x_{24}^4 x_{13}^2 h(2,3,1;2,4,4)$, which does not have an interpretation in terms of two-loop four-particle planar momentum integrals.
The same `bad' integral comes from $h(1,2,5;1,3,4)$ and $p(1;2,4;3,5)$ and their
cyclic images. We find that $b(1,2,3,4)$ cancels in the sum of all terms in the right-hand side of \re{fin} as $x_4\to x_5$, leading to
\begin{align}\label{FG-s}
\lim_{x_4\to x_5} F_G^{(2)}  &= 2\, x_{13}^4 x_{24}^2 h(1,3,2;1,3,4)+ 2\, x_{24}^4 x_{13}^2 h(2,4,1;2,4,3)
 -  [x_{13}^2 x_{24}^2 g(1,2,3,4)]^2\,.
\end{align}
Again, we verify that the expression in the right-hand side coincides with the
ratio of four-point correlators \p{G-soft}.

Comparing \re{FA-s} and \re{FG-s}, we conclude that the two-loop corrections to $F_A$ and $F_G$ coincide in the soft limit $x_4\to x_5$,
\begin{align}\label{dual-s}
\lim_{x_4\to x_5} \left( F_A^{(2)} - F_G^{(2)} \right) = 0\,.
\end{align}
In the next subsection we argue that for arbitrary $x_i$ the difference $F_A^{(2)} - F_G^{(2)}$ is a constant. Together with \re{dual-s} this immediately implies that $F_A^{(2)} = F_G^{(2)}$, as announced in \p{dual2}.

\subsubsection{Conformal symmetry and proof of the integral identity}

As was already mentioned, the duality relation  $F_A^{(2)}=F_G^{(2)}$
implies an identity between two-loop integrals of various topologies.  Evaluating the difference $F_G^{(2)}-F_A^{(2)}$ with the help of \re{fin} and \re{amp2} and equating it to zero, we formulate the identity that we expect to find as follows:
\begin{align}\notag\label{magic}
0& = \, x^2_{13} \, x^2_{24} \, \left[
 x^2_{13} \, h(1,2,3;1,3,4) \, +   x^2_{24} \, h(1,2,4;2,3,4) - \, x^2_{14} \, h(1,2,4;1,3,4) \right]
\\[2mm]  \notag&
+ \, x^2_{13} \, x^2_{14} \, x^2_{25} \left[ 2 \, h(1,2,5;1,3,4) \, - \,
h(1,2,3;1,4,5) \, - \, h(1,2,4;1,3,5) \right]
\\[2mm] & \notag
+ \, x^2_{24} \, x^2_{35} \left[  x^2_{25} \, p(1;2,5;3,4) \, - \,
x^2_{24} \, p(1;2,4;3,5) \, - \, x^2_{35} \, p(1;3,5;2,4) \right]
\\[2mm] 
&
+ \ft1{2}
x_{13}^2 x_{24}^2 g(1,2,3,4) x_{35}^2 x_{14}^2   g(1,3,4,5)
+ \text{(cyclic)}\,.
\end{align}
In the previous subsection we showed that this relation holds in the soft limit.

The proof of \re{magic} goes as follows. The Feynman integrals in
\re{magic} are regularized dimensionally with $D=4-2\ep$. We will
first show that the sum of Feynman integrals in the right-hand
side of \re{magic} is finite as $\ep\to 0$. This will allow us to
remove the regulator, i.e. to restore the $D=4$ integration
measure. Since all integrals we are dealing with are
pseudo-conformal, the expression in the right-hand side of
\re{magic} will thus become an exactly conformally invariant
function of $x_1,\ldots,x_5$. Since one cannot construct conformal
cross-ratios from the five external points with light-like
separated neighbors $x_{i,i+1}^2=0$,  conformal invariance will
imply that  the right-hand side of \re{magic} is a constant. We
have already seen that the right-hand side of  \re{magic} vanishes
in the soft limit, so the constant must be zero.

The direct way to prove finiteness is to evaluate the pole part of
(\ref{magic}) by the Mellin-Barnes (MB) method (see, e.g., \cite{mbMethod})
using the package \cite{czakon} for the $\epsilon$-expansion and
evaluation of the representations. At $O(1/\epsilon^4)$ through
$O(1/\epsilon^2)$ the program finds only one-parameter integrals
which we have analytically evaluated. At $1/\epsilon^4$ and
$1/\epsilon^3$ one has to show the cancelation of rational
numbers and simple logarithms, respectively, which was in either
case immediate. At $1/\epsilon^2$ we needed Landen's identity on
dilogarithms, but once again the proof was ultimately simple. With
our MB representations, the highest integrals contributing to the
simple pole in $\epsilon$ were of dimension four (from the
pentabox). The precision of the numerical evaluation used by the
package \cite{czakon} was therefore good: For all kinematic points
in our (rather general) sample set we obtained values like
$0.00(2)/\epsilon$. Within the given accuracy the sum of integrals
is seen to be finite.\footnote{An analytical proof of the absence of the
simple pole, too, should be feasible given the relatively low
dimensionality of the MB-integrals.} {As we have just explained, the finiteness of the integral implies that it equals zero since
a non-vanishing constant part is ruled out by the soft limit.}

Alternatively, we can show finiteness {analytically} by combining the sum of integrals in \re{magic} into a single integral
of the form
\begin{align}\label{P-rep}
\text{r.h.s. of Eq.\re{magic}} = \int \frac{d^D x_0 d^D x_{0'}\, P(x_0,x_{0'};x_i)}{(x_{10}^2\ldots x_{50}^2 )(x_{1{0'}}^2\ldots x_{5{0'}}^2) x_{0{0'}}^2}\,,
\end{align}
where, by construction, the polynomial $P(x_0,x_{0'},x_i)$ is invariant under cyclic shifts of the external points $x_i$ (with $i=1,\ldots,5$)  and  under the exchange of the integration points $x_0\leftrightarrow x_{0'}$. Each integral in \re{magic} gives a contribution to $P(x_0,x_{0'},x_i)$ in the form of a product
of seven distances $x_{ij}^2$ with various choices of indices $i$ and $j$.
For instance, the third $h-$integral in the first line of \re{magic} produces
the contribution $ (- x_{13}^2 x_{14}^2 x_{24}^2 x_{30}^2x_{50}^2 x_{2{0'}}^2 x_{5{0'}}^2)$, while the contribution of the cross-product of one-loop integrals in the last
line of \re{magic} looks as $(-\frac12 x_{13}^2 x_{24}^2 x_{14}^2 x_{35}^2 x_{50}^2 x_{2{0'}}^2x_{0{0'}}^2$). To save space, we do not present the explicit expression for
$P(x_0,x_{0'};x_i)$.
Due to the pseudo-conformal property of the integrals in \re{magic}, the polynomial $P(x_0,x_{0'},x_i)$ is covariant under conformal
transformations, with weight 4 at points $x_0$ and $x_{0'}$. Given the weight $(-8)$ of the denominator, and if the integration measures can be made four-dimensional, the integral will be conformally covariant, as stated above.

Let us identify the potential divergences of the integral \re{P-rep}. They could only come from the part of the phase space of $x_0$ and $x_{0'}$, in which some of the propagators in the denominator of  \re{P-rep} vanish simultaneously.
Notice that upon the change of variables, $p_1=x_{12},\ldots,p_5=x_{51}$ (with $p_i^2=0$) supplemented with $k=x_{10}$ and $k'=x_{1{0'}}$, the integral \re{P-rep} can be rewritten as a conventional two-loop momentum-space Feynman integral with $k$ and $k'$ being the loop momenta and $p_1,\ldots,p_5$ defining the external leg
momenta. It is well known that infrared divergences originate from integration over the loop momenta collinear to the light-like momenta of the external legs. For instance, suppose that
$k^\mu$ is collinear either to $p_1^\mu$ and to $p_2^\mu$. To
analyze the divergences, it is convenient to employ the so-called Sudakov decomposition of the loop momentum,
\begin{align}
 k^\mu \equiv x_{10}^\mu= \alpha p_1^\mu +\beta p_2^\mu + k_\perp^\mu\,,
\end{align}
where $\alpha,\beta$ are scalar variables and $k_\perp^\mu$ are two-dimensional transverse momenta, $(k_\perp p_1)=(k_\perp p_2)=0$.
In terms of Sudakov's variables, the $x_0-$integral in \re{P-rep} takes the form
\begin{align}\label{Sud}
\int \frac{d^4 x_0}{x_{01}^2 x_{20}^2 x_{30}^2}\big[ \ldots\big]  = \int  \frac{x_{13}^2\, d\alpha d\beta d^{2} k_\perp\, \big[ \ldots\big]}{(x_{13}^2\alpha\beta - k_\perp^2)(x_{13}^2(1-\alpha)\beta + k_\perp^2)(x_{13}^2\alpha(1-\beta) + k_\perp^2)}\,,
\end{align}
where $[\ldots]$ denotes the remaining terms in the right-hand side of \re{P-rep}.

It is easy to see that the integral \re{Sud} develops logarithmic divergences originating from the integration region
\begin{align}\label{reg-IR}
x_{13}^2\alpha\beta - k_\perp^2 = O(\rho^2)\,,\qquad k_\perp = O(\rho)
\end{align}
with $\rho\to 0$. Depending on the hierarchy between $\alpha$ and $\beta$ we can
distinguish three subregions: For $\alpha,\beta=O(\rho)$ we have $k^\mu=O(\rho)$, for $\alpha=O(\rho^0)$, $\beta=O(\rho^2)$ we have $k^\mu=\alpha p_1^\mu + O(\rho)$ and, finally, for $\alpha=O(\rho^2)$, $\beta=O(\rho^0)$ we have $k^\mu=\beta p_2^\mu + O(\rho)$. In terms of  the dual $x-$variables, this corresponds to the limit where the integration point $x_0$ approaches  either the external point,
$x_0\to x_1$, or one of  the light-like segments $[x_1,x_2]$ and $[x_2,x_3]$, respectively. Due to the cyclic symmetry of the integral \re{P-rep}, divergences are also produced when $x_0$ approaches the other cusp points $x_i$ and  the light-like segments $[x_i,x_{i+1}]$. The same analysis applies to the integral with respect to $x_{0'}$.

Then, for the integral \re{Sud} to be finite, the expression inside $[\ldots]$
should be finite and, in addition, it should vanish in the region \re{reg-IR}. This leads in its turn to the condition for the
polynomial $P(x_0,x_{0'},x_i)$ in the right-hand side of \re{P-rep} to vanish sufficiently fast when $x_0$ and/or $x_{0'}$ approach one the potentially dangerous regions explained above. We return to \re{P-rep} and verify that the polynomial $P(x_0,x_{0'};x_i)$ indeed satisfies this condition. Thus, the integral \re{P-rep} remains finite as $\epsilon\to 0$. Together with conformal invariance this immediately leads to the identity \re{magic}.

\section{Conclusions}

The main result of this paper is the observation that computing the loop corrections to the correlators of protected operators by the Lagrangian insertion method, and taking the light-cone limit, gives us the
integrands of the MHV gluon scattering amplitudes in the dual momentum space. These integrands can be evaluated directly in four
dimensions since they are finite and  explicitly (dual) conformally covariant. The divergences appear in
the integrals over the insertions points for the correlator (UV divergences) or, equivalently, over the loop momenta for the amplitude (IR divergences). In order to achieve an exact matching of the two objects, we need to dimensionally regularize the theory, not as one would do for a correlator computation in coordinate space, but in the way that is natural for the dual theory in momentum space. We call this dual infrared regularization.

At present we have no understanding of the field-theory mechanism responsible for this surprising relation between two seemingly very different objects. In fact, the surprise is not total, since we already know that MHV amplitudes are dual to light-like Wilson loops. The latter, as shown in \cite{AEKMS}, are intimately related to correlators in the light-cone limit.

Further tests are needed to confirm our conjecture. In particular, a crucial test will be the calculation of the six-point correlator at two loops and its comparison to the six-gluon amplitude. This is similar to the test of the Wilson loop/amplitude duality at six points and two loops \cite{Drummond:2008aq,Bern:2008ap}. Before it, one could suspect that the matching of the four- and five-point objects was simply due to (dual) conformal symmetry, which fixes their form completely. As we saw in our five-point two-loop correlator computation in Sect.~\ref{52loop}, conformal symmetry plays an important role here too. However, at six points it will not be sufficient to explain the matching, if it is confirmed.

The Lagrangian insertion procedure described here can give us a simple way to compute the integrands that determine the higher-loop corrections to amplitudes. We would like to emphasize that our one- and two-loop correlator computations involve a handful of Feynman graphs. This is in striking contrast to the huge number of graphs needed, if one would attempt to do a straightforward amplitude calculation (not using unitarity methods).

The fact that the integrand of the amplitude can be obtained from a tree-level computation raises the hope that one might be able to profit from some hidden enhanced symmetry of the tree-level correlator. This is motivated by the analogy with the tree-level scattering superamplitudes, which are known to have a dynamical symmetry called dual conformal symmetry \cite{Drummond:2008vq}. If something similar exists also for the correlators, it might eventually lead us to the integrability of the loop amplitudes.

A related issue is the observation that the correlators we have discussed enjoy the full superconformal symmetry of the $\cN=4$ SYM theory. In this paper we have explored the lowest scalar components of the supersymmetric correlators, and we have shown that they match the gluon MHV amplitudes. The question arises if the correlator defined in superspace has some relation to the superamplitude, and thus possibly to non-MHV amplitudes.

\section*{Acknowledgments}

{We are grateful to Juan Maldacena for carefully reading the manuscript and making many useful suggestions.} We would like to thank Fernando Alday, Babis Anastasiou, Vladimir Braun, Evgueny Ivanov, Henrik Johansson, David Kosower, Radu Roiban, Misha Shifman, Boris Zupnik  for interesting discussions. For BE and ES it is a pleasure to acknowledge the collaboration with Christian Schubert, with whom part of the supergraph method from Appendix~\ref{ApB} was developed. GK is grateful to Vladimir Bazhanov  and to the Australian National University, Canberra,
and ES is grateful to Nima Arkani-Hamed and the Institute for Advanced Study, for warm hospitality at various stages of this work.  This work was supported in part by the French Agence Nationale de la Recherche under grant
ANR-06-BLAN-0142.

\appendix

\section*{Appendices}

\section{Lagrangian insertion procedure in  ${\cal N}=2$ harmonic superspace}\label{ApB}

In this appendix we give a very brief overview of harmonic superspace, in particular of the Feynman rules we need here. More details can be found in \cite{hh,Galperin:2001uw,hsgr}. We then summarize the Lagrangian insertion procedure, developed in \cite{Howe:2000hz,Eden:2000mv} for the case of four-point correlators. Finally, we explain how to adapt this procedure to five (and more) points.

\subsection{${\cal N}=4$ SYM in $\cN=2$ terms}

The basic ingredients of the  ${\cal N}=4$ SYM theory in $\cN=2$ terms are the hypermultiplet and the super-Yang-Mills (SYM, or vector) multiplet. The ${\cal N}=4$ SYM action consists of two terms:
\begin{equation}\label{N4sym'}
  S_{\mbox{\scriptsize {\cal N}=4 \rm SYM}} =  S_{\mbox{\scriptsize {\cal N}=2 \rm  SYM}} +
 S_{\mbox{\scriptsize \rm HM}}\;.
\end{equation}
Below we give a short description of each multiplet and its action.

\subsubsection{${\cal N}=2$ hypermultiplet in harmonic superspace}

The ${\cal N}=2$ massless matter (or hyper)multiplet consists of an R-symmetry $SU(2)$ doublet of complex scalars $\phi^i(x)$ (with $i=1,2$) and of two Majorana spinors and $SU(2)$ sunglets $\psi_\a(x)$, $\bar\psi_\da(x)$ and $\kappa_\a(x)$, $\bar\kappa_\da(x)$. Their supersymmetry transformations close only on shell. Going off shell requires the introduction of an infinite set of auxiliary fields \cite{hh}. This is achieved by extending the space-time by  two extra compact dimensions in the form of a sphere $S^2$. The latter is described in terms of {\it harmonic
variables}  $u^{\pm i}$ which form a matrix of $SU(2)$,
\begin{equation}\label{defharvar}
  \parallel u \parallel\ \in {SU}(2):\qquad  u^{+i}u^-_i = 1\,,\qquad
\overline{u^{+i}} = u^-_i \equiv \epsilon_{ij} u^{-j}\,, \qquad \epsilon_{12} = - \epsilon^{12} = 1\,,
\end{equation}
and parametrise the sphere $S^2\sim SU(2)/U(1)$. A harmonic function $f^{(q)}(u^\pm)$
of $U(1)$ charge $q$ is a function of $u^{\pm i}$ invariant under
the action of the group $SU(2)$ (which rotates the index $i$ of
$u^{\pm i}$) and homogeneous of degree $q$ under the action of the
group $U(1)$ (which rotates the index $\pm$ of $u^{\pm i}$). Such
functions have infinite harmonic expansions on $S^2$ whose
coefficients are $SU(2)$ tensors (multispinors).

In this framework the hypermultiplet is described by a harmonic
superfield $q^+(x,\q,\bq,u)$ of $U(1)$ charge $+1$ satisfying the Grassmann (or G-)analyticity constraints
\begin{equation}
D^+_\alpha q^+ = \bar D^+_{\dot\alpha} q^+ = 0\,, \label{5.2.5a}
\end{equation}
where
\begin{equation}
D^+_\alpha = D^i_\alpha u^+_i\,, \qquad \bar D^+_{\dot\alpha} = \bar
D^i_{\dot\alpha}u^+_i
\label{5.2.5b}
\end{equation}
and $D^i_\alpha$, $\bar D^i_{\dot\alpha}$ are the usual
supersymmetric spinor derivatives. These constraints can be solved
explicitly in the G-analytic basis in superspace \be\label{ganba}
x^{\alpha\dot\alpha}_A = x^{\alpha\dot\alpha} - 4i\theta^{\alpha
(i} \bar\theta^{\dot\alpha j)} u^+_i u^-_j\;, \quad
\theta^\pm_{\alpha,\dot\alpha} =
u^\pm_i\theta^i_{\alpha,\dot\alpha} \end{equation} where
$x^{\alpha\dot\alpha} = x^\mu \sigma_\mu^{\alpha\dot\alpha}$ and
$(ij)$ means weighted symmetrization. In this basis $q^+$ becomes
a function of $\theta^+, \bar\theta^+$ only, i.e., a G-analytic
superfield $q^+(x_A,\theta^+,\bar\theta^+,u)$.

As mentioned earlier, the ${\cal N}=2$ supermultiplet can exist
off shell because an infinite number of auxiliary fields
(coming from the harmonic expansion on $S^2$) are present. On
shell these  auxiliary fields are eliminated by the harmonic
(or H-)analyticity condition (equation of motion)
\begin{equation}\label{EMo}
  D^{++}q^+ = 0\;.
\end{equation}
Here $D^{++}$ is the harmonic derivative on $S^2$ (the raising
operator of the group $SU(2)$ realized on the $U(1)$ charges,
$D^{++}u^+=0,\; D^{++}u^-=u^+$). In the G-analytic basis (\ref{ganba}) it
becomes a supercovariant operator involving space-time derivatives:
\begin{equation}\label{D++}
  D^{++} = u^{+i}{\partial\over\partial u^{-i}}
-4i\theta^{+\alpha}\bar\theta^{+\dot\alpha} {\partial\over\partial
x^{\alpha\dot\alpha}_A} \;.
\end{equation}
It is then easy to show that the free on-shell hypermultiplet becomes an ``ultrashort" superfield:
\begin{eqnarray}
q^+(x_A,\q^+, \bq^+,u) &=& \phi^i(x_A)u^+_i
+\theta^{+\alpha}\psi_\alpha(x_A)
+\bar\theta^+_{\dot\alpha}\bar\kappa^{\dot\alpha}(x_A)+
4i\q^+\sigma^\mu\bq^+ \pa_\mu \phi^i(x_A) u^-_i\,,
\label{onshq'}
\end{eqnarray}
where the physical scalars $\phi^i$
and spinors $\psi_\alpha,\ \bar\kappa^{\dot\alpha}$ satisfy their
massless field equations $ \square \phi^i(x)=\dslash\psi
=\dslash\bar\kappa=0\,.$

The equation of motion (\ref{EMo}) can be derived from an action given
by an integral over the G-analytic superspace:
\begin{equation}\label{6.7.1}
S_{\mbox{\scriptsize HM}} = -2 \int
dud^4x_Ad^2\theta^+d^2\bar\theta^+\; \mathrm{Tr}\left(\tilde
q^{+}D^{++}q^+\right)\;.
\end{equation}
Here $\tilde q^+(x_A,\theta^+,\bar\theta^+,u)$ is the conjugate of $q^+$. The conjugation
$\ \widetilde{}\ $ combines usual complex conjugation with the antipodal map on $S^2$ in a way to preserve G-analyticity.  This action is real (with respect to the $\ \widetilde{}\ $ conjugation)
which can be seen by integrating $D^{++}$ by parts. In this sense
the action (\ref{6.7.1}) resembles the Dirac action for fermions,
although the superfield $q^+$ is bosonic.

\subsubsection{${\cal N}=2$ SYM multiplet in harmonic superspace}

The ${\cal N}=2$ SYM gauge potential is introduced by covariantizing the action
(\ref{6.7.1}) with respect to a Yang-Mills group with G-analytic
parameters $\lambda(x_A,\theta^+,\bar\theta^+,u)$. To this end one
replaces the harmonic derivative in (\ref{6.7.1}) by the following
covariant one:
\begin{equation}\label{covhder}
D^{++}\rightarrow  D^{++} +
igV^{++}(x_A,\theta^+,\bar\theta^+,u)\,,
\end{equation}
where $g$ is the gauge coupling constant.
The gauge potential is
described by a real ($\widetilde {V}^{++} =V^{++}$) G-analytic
superfield of charge $+2$ (equal to the charge of $D^{++}$). The
matter and gauge superfields are subject to the usual gauge
transformations:
\begin{equation}\label{6.7.5}
{q^+}' = e^{ig\lambda}q^+\,, \ \ \ {V^{++}}' = -{i\over
g}e^{ig\lambda}D^{++}e^{-ig\lambda} +
e^{ig\lambda}V^{++}e^{-ig\lambda}\;,
\end{equation}
so that the covariantized action (\ref{6.7.1})
\begin{equation}
  S_{\mbox{\scriptsize \rm HM/SYM}}
    = - 2 \int
dud^4x_Ad^2\theta^+d^2\bar\theta^+\ {\rm Tr}( \tilde q^{+}D^{++}q^+ + ig\, \tilde q^{+}V^{++}q^+) \label{HMcov}
\end{equation}
is indeed gauge invariant.

In the non-supersymmetric Wess-Zumino gauge the gauge potential has the component expansion
\begin{eqnarray}\label{6.10.1}
 V^{++}_{\rm WZ}(x_A,\q^+,\bq^+,u) &=& -2i\theta^+\sigma^\mu\bar\theta^+
A_\mu(x_A) - i \sqrt2 (\theta^+)^2\bar \varphi(x_A) + i\sqrt2(\bar\theta^+)^2 \varphi(x_A)
 \\[2mm] && +\,4(\bar\theta^+)^2\theta^{+\alpha}\lambda^i_{\alpha}(x_A) u^{-}_i
-4(\theta^+)^2\bar\theta^+_{\dot\alpha}\bar\lambda^{\dot\alpha i }(x_A) u^{-}_i
 +\, 3(\theta^+)^2(\bar\theta^+)^2 Y^{ij}(x_A)u^-_iu^-_j\,, \nn
\end{eqnarray}
containing the fields of the $\cN=2$ {\it off-shell} vector multiplet: the gauge field $A_\mu$, the complex physical scalar $\varphi$, the doublet of Majorana gluinos $\lambda^i_{\alpha}$, $\bar\lambda^{\dot\alpha i }$ and the triplet of real auxiliary fields  $Y^{ij}$.

The gauge invariant action for $V^{++}$ can be written down either directly in terms of $V^{++}$ \cite{hsgr,Galperin:2001uw}, or in terms of the chiral  superfield strength $W(x_L,\theta^{i\alpha})$ (or its conjugate antichiral $\bar W(x_R,\bar\theta_{i\dot\alpha})$):
\begin{equation}\label{SYMact}
  S_{\mbox{\scriptsize {\cal N}=2 \rm SYM}} =
{1\over  2 g^2}\int d^4x_L d^4\theta\;  {\rm Tr}\;W^2 = {1\over 2
g^2}\int d^4x_R d^4\bar\theta\;  {\rm Tr}\;\bar W^2\;,
\end{equation}
where
\begin{equation}\label{chiba}
    x^{\alpha\dot\alpha}_L = x^{\alpha\dot\alpha} - 2i\theta^{i\alpha} \bq^\da_i
\end{equation}
are the space-time coordinates  in the chiral basis, and $x_R = {\bar x_L}$ are the antichiral ones.
In a
topologically trivial background these two forms are equivalent (up
to a total space-time derivative), due to the Bianchi identity $ D^{i\a} D^j_\a W= \bar D^i_{\dot\alpha}\bar D^{j\dot\alpha} \bar W$.

Unlike the G-analytic potential $V^{++}$, the field strength $W(x_L,\q)$ is a chiral
superfield which  does not depend on the harmonic variable $u^\pm$. It can be
expressed as a power series in $V^{++}$, involving multiple harmonic integrals  \cite{Zup}\begin{equation}\label{WV}
  W = {i\over 4} u^+_iu^+_j \bar D^i_{\dot\alpha}\bar D^{j\dot\alpha}
\sum^\infty_{r=1} \int du_1\ldots du_r\; {(-ig)^{r} V^{++}(u_1)
\ldots V^{++}(u_r) \over (u^+u^+_1)(u^+_1u^+_2) \ldots (u^+_ru^+)}\;,
\end{equation}
where $(u^+u^+_1) \equiv u^{+i} \ep_{ij} u^{+j}_1$, etc. In terms of component fields we have (in the Abelian case)
\begin{equation}\label{WV'}
  W = \varphi(x) + \theta^{\alpha i} \lambda_{\alpha i}(x) + \theta^{\alpha i}\theta^{\beta j}(\epsilon_{ij} (\sigma^{\mu\nu})_{\alpha\beta} F_{\mu\nu}(x) + \epsilon_{\alpha\beta} Y_{ij}) +i (\q^3)_{i}^\a \pa_{\a\da} \bar\lambda^{\da i}(x) + \q^4 \square \bar\varphi(x)\,.
\end{equation}

With the hypermultiplet matter in the adjoint
representation of the gauge group, the two actions (\ref{HMcov})
and (\ref{SYMact}) describe the ${\cal N}=4$ SYM theory,
\begin{equation}\label{N4sym}
  S_{\mbox{\scriptsize {\cal N}=4 \rm SYM}} =  S_{\mbox{\scriptsize {\cal N}=2 \rm SYM}} +
 S_{\mbox{\scriptsize \rm HM/SYM}}\;.
\end{equation}

As mentioned earlier, the main advantage of the ${\cal N}=2$ harmonic superspace formulation
is the possibility to quantize the theory in a straightforward and manifestly $\cN=2$ supersymmetric way \cite{hsgr}. Further, compared to the ${\cal N}=1$ chiral matter superfields,  the  ${\cal N}=2$ hypermultiplet composite operators $O={\rm Tr} (q^+)^k$, etc., need no covariantization, hence no presence of the gauge superfield in the definition of the correlators $\vev{O\ldots \tilde O}$. The hypermultiplet matter interacts with the gauge sector only through a single trilinear vertex, which considerably simplifies the Feynman diagrams. The true non-Abelian nature of the theory is encoded in the gauge self-interactions (as well as in the ghost sector, but we do not need it here).

\subsection{Feynman rules}\label{FR}

In this section we give a subset of the Feynman rules for the combined ${\cal N}=2$ matter$+$gauge system which are sufficient for the one- and two-loop calculations we do in this paper (the complete set can be found in \cite{hsgr,Galperin:2001uw}).

The hypermultiplet propagator is the solution to the Green's function equation
\begin{equation}\label{hypermultipletproeq}
  D^{++}_1\langle \tilde q^+(1) q^+(2)\rangle = \delta^4(x_{A1} - x_{A2})\delta^2(\q^+_1-\q^+_2)\delta^2(\bq^+_1-\bq^+_2)\delta(u_1,u_2)\,,
\end{equation}
and is given by
\begin{equation}\label{completeqprop}
  \langle \tilde q^+_a(1) q^+_b(2)\rangle =
 {(12)\over 4\pi^2\hat x^2_{12}}\delta_{ab} =\
\psfrag{1}[cc][cc]{$\scriptstyle 1,a$}\psfrag{2}[cc][cc]{$\scriptstyle 2,b$}
  \includegraphics[width=0.15\textwidth]{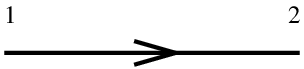}
\,.
\end{equation}
Here $a,b$ are color indices  and
\begin{equation}\label{def(12)}
  (12) = -(21) = u^{+i}_1 \epsilon_{ij} u^{+j}_{2}
\end{equation}
is a shorthand for the $SU(2)$ invariant but $U(1)$ covariant contraction of the two
harmonics. The coordinate difference
\begin{equation}\label{hataa}
  \hat x^\mu_{12} = x^\mu_{A1}
- x^\mu_{A2} + {2i\over (12)}[(1^-2) \theta^+_1\sigma^\mu \bar\theta^+_1 +
(2^-1) \theta^+_2\sigma^\mu \bar\theta^+_2 +
\theta^+_1 \sigma^\mu\bar\theta^+_2 +
\theta^+_2\sigma^\mu \bar\theta^+_1]\,,
\end{equation}
where, e.g., $(1^-2) = u^{-i}_1 \epsilon_{ij} u^{+j}_{2}$,
is invariant under the Poincar\'e supersymmetry transformations in the G-analytic basis \p{ganba}:\footnote{ To check this one makes use of the harmonic cyclic identity
$
  (1^-2)1 + (21)1^- + (11^-)2 = 0
$
and of the defining property $(11^-)=1$ (see (\ref{defharvar})).}
\begin{equation}\label{Qsusy}
    \delta_Q x^{\alpha\dot\alpha}_A =
-4iu^-_i(\epsilon^{i\alpha}\bar\theta^{+\dot\alpha} +
\theta^{+\alpha}\bar\epsilon^{i\dot\alpha})\,, \qquad \delta_Q \theta^{+\alpha, \dot\alpha} = u^+_i\epsilon^{i\alpha,
\dot\alpha}\,, \qquad \delta_Q u^\pm_i = 0\,.
\end{equation}

Setting the Grassmann variables in (\ref{completeqprop}) to zero, we find the propagator for the physical scalars $\phi^i_a(x)$ projected with harmonics (recall (\ref{onshq'})):
\begin{equation}\label{qproat0}
  \langle \tilde q^+_a(1) q^+_b(2)\rangle_{\theta=0} = \langle u^{+i}_{1} \bar \phi_{i a}(x_1) | u^+_{2j} \phi^j_b(x_2)  \rangle  = \frac{(12)}{4\pi^2 x^2_{12}} \delta_{ab}
  \;.
\end{equation}

The gauge field (gluon) propagator depends on the gauge we have chosen. In our loop calculations we will only need the propagator
$\langle W(1)V^{++}(2)\rangle$, having one
chiral end (the field strength $W(x_{L1},\theta_1)$) and one
G-analytic end (the SYM potential
$V^{++}(x_{A2},\theta^+_{2},\bar\theta^+_{2},u_2)$). It is independent of the gauge and has the following form
\begin{equation}\label{propWV'}
  \langle W_a(1)V_b^{++}(2)\rangle =  - {g \delta_{ab}\over
2\pi^2\widetilde x^2_{12}}
(\theta_{12})^2 = \
\psfrag{1}[cc][cc]{$\scriptstyle 1,a$}\psfrag{2}[cc][cc]{$\scriptstyle 2,b$}\vspace*{2mm}
  \includegraphics[width=0.2\textwidth]{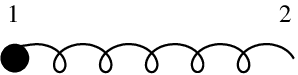}
\end{equation}
It involves the coordinate differences
\begin{equation}\label{chhat}
    \widetilde x^{\alpha\dot\alpha}_{12} = x^{\alpha\dot\alpha}_{L1}
- x^{\alpha\dot\alpha}_{A2} - 4iu^-_{2i}\theta^{i\alpha}_1\;
\bar\theta^{+\dot\alpha}_2\,, \qquad \theta^{\alpha}_{12} =  u^+_{2i}\theta^{i\alpha}_1 - \theta^{+\alpha}_{2}
\end{equation}
and is invariant under the Poincar\'e supersymmetry transformations in the chiral basis for $x_{L1}$,
\begin{equation}\label{trafosusy}
  \delta_Q x^{\alpha\dot\alpha}_L = -4i\theta^{\alpha
i}\bar\epsilon^{\dot\alpha}_i\;, \qquad \delta_Q \theta^{i\alpha} = \epsilon^{i\alpha}  \,,
\end{equation}
and \p{Qsusy} in the G-analytic basis for $x_{A2}$.

Finally, the only interaction vertex we shall need here is the gluon-to-matter coupling which can be read off from the covariantized hypermultiplet action \p{HMcov}:

\begin{align}\label{vertex}
\parbox[c]{0.2\textwidth}{\psfrag{0}[cc][cc]{$0$}
\psfrag{q}[cc][cc]{$q^+_c$}
\psfrag{qt}[cc][cc]{$\tilde q^+_a$}
\psfrag{V}[cc][cc]{$V_b^{++}$}
  \includegraphics[width=0.2\textwidth]{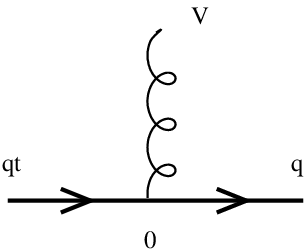}} \qquad  = \qquad  ig f_{bac}\int d^4x_{0} du_0 d^4\theta_{0}^+
\end{align}
The SYM action (\ref{SYMact}), \p{WV} is non-polynomial in $V^{++}$, so it contains infinitely many vertices (but, of course, becomes polynomial in the Wess-Zumino gauge \p{6.10.1}). At the one- and two-loop levels that we are studying in this paper only the cubic non-Abelian vertex can appear, but later on we shall see that all such graphs are irrelevant.

\subsection{The insertion procedure: Four points}\label{ip4p}

Here we illustrate the ${\cal N} =2$ Lagrangian insertion procedure
described in Sect.~\ref{Cpos}, on the example of the two-loop
four-point correlator $G_4$ (\ref{44pt}) of hypermultiplet bilinears
${\cal O}=O_{\theta^+=\bar \theta^+=0}$ with ${O}= {\rm Tr}(q^+)^2$ and
${\tilde O}= {\rm Tr}(\tilde q^+)^2$. In particular, following
\cite{Eden:2000mv} we explain the important role of ${\cal N}=2$
superconformal symmetry, which allows us to drastically simplify
the two-loop calculations.

The two-loop (order $g^4$) corrections to the four-point
correlator $G_4$ can be obtained by a double insertion of the
${\cal N}=2$ SYM chiral action
\begin{equation} \label{n2sym}
S_{{\cal N}=2~\mbox{\tiny \rm SYM}}=\int \rmd^4 x\rmd^4\theta~ {L}(x_L,\theta)\, ,
~~~~{L}=\frac{1}{2 g^2}\mbox{Tr}(W(x_L,\theta))^2 \, .
\end{equation}
To see this, we first rescale the gauge potential
$V^{++}\ \to \ g^{-1} V^{++}$, with the effect that the coupling only
appears in front of the SYM action (\ref{SYMact}), but not inside the
field strength $W$ (\ref{WV}). It also drops out of the gauge/matter
coupling (\ref{HMcov}), thus (\ref{vertex}) loses the explicit $g$.
As another consequence, the gauge propagator (i.e. the
inverse of the gauge kinetic term) is scaled up by $g^2$, which introduces a factor of $g^2$ in the right-hand side of (\ref{propWV'}).

Let us write out the perturbative expansion of $G_4$:
\begin{equation}
G_4 \, = \, G_4^{(0)} + g^2 \, G_4^{(1)} + \, g^4 \, G_4^{(2)} + \ldots\ ,
\end{equation}
so that
\begin{equation}
G_4^{(2)} \, = \, \frac{1}{2} \,
\left(\frac{\pa}{\pa g^2}\right)_{g=0}^2 \, G_4 \, .
\end{equation}
On the other hand, by considering the effect of the differentiation on the
path integral we find
\begin{eqnarray} \label{ins}
\frac{1}{2} \, g^4 \left(\frac{\pa}{\pa g^2}\right)^2 G_4 & = &
i \int \rmd^4 x_0\rmd^4\theta_0\ \langle {L}(0) {O}(1) \tilde{O}(2) {O}(3)
\tilde{O}(4)\rangle \\ && - \frac{1}{2} \int \rmd^4 x_0\rmd^4\theta_0\
\rmd^4 x_{0'}\rmd^4\theta_{0'} ~\langle {L}(0){L}({0'}) {O}(1)
\tilde{O}(2) {O}(3) \tilde{O}(4)\rangle \, . \nonumber
\end{eqnarray}
The left-hand side in the last formula starts at $O(g^4)$, which leads to a
puzzle: The perturbative expansion of the first term on the right-hand side starts at
$O(g^2)$, while the second term seemingly starts at $O(g^4)$. The only way to
produce a compensating $O(g^2)$ contribution from there is to insert both
${L}(0),{L}({0'})$ into one gluon line. This means inserting the
chiral-to-chiral
propagator (it contains just the propagator of the auxiliary field $Y$ in \p{WV'})
\begin{equation}
\langle W(0) W({0'})\rangle =2ig^2 \delta^4(x_{L 0} - x_{L 0'})
\delta^4(\theta_{0} - \theta_{0'})
\end{equation}
into that gluon line. Upon performing
the chiral superspace integration over point ${0'}$, the $O(g^2)$ contribution
from the single insertion term is identically canceled. This remains true in
general: $00'$ contact terms from the double insertion term identically
cancel the single insertion term. Returning to $O(g^4)$:
\begin{equation}
G_4^{(2)} \, = \, - \frac{1}{2 g^4} \int \rmd^4 x_0\rmd^4\theta_0\
\rmd^4 x_{0'}\rmd^4\theta_{0'} ~\langle {L}(0){L}({0'}) {O}(1)
\tilde{O}(2) {O}(3) \tilde{O}(4)\rangle_{g^4}^\mathrm{reg}
\end{equation}
where the superscript indicates that the $00'$ contact terms are to be omitted.
For the class of graphs we find below this simply means not to contract
$W(0)$ and $W({0'})$. We stress that this $(4+2)-$point correlator is once again
at Born level; it comes with $g^4$ because there are four Yang-Mills propagators
and two explicit factors $g^{-2}$ from the Lagrangian insertions. Below
we show that this correlator with insertions has to be nilpotent,
i.e. proportional to $\q^8$.

\subsubsection{Structure of the nilpotent superconformal covariant}\label{snsc}

The most important feature of the new six-point correlator is its superconformal symmetry. Indeed, it involves gauge-invariant composite operators, ${O}$, ${\tilde O}$ and ${L}$. As explained in Sect.~\ref{Cpos}, all of these operators are particular projections of the $\cN=4$ half-BPS protected operator in the $\mathbf{20'}$ of $SU(4)$. As such, they need no renormalization and have well-defined superconformal properties. As long as we keep the end points in this correlator apart, nothing can break the $\cN=2$  superconformal symmetry of the theory.  This imposes rather strong constraints on the general form of the correlator.

The $\cN=2$ superconformal algebra has an $SU(2)\times U(1)$ automorphism group (R symmetry). The $U(1)$ factor (to be distinguished form the harmonic $U(1) \subset SU(2)$) acts only on the odd  superspace variables, $R[\q]=1/2$, $R[\bq] = -1/2$. From the SYM action \p{n2sym} we deduce $R[{L}]=2$ and $R[W]=1$. At the same time, the hypermultiplets $q^+$ and $\tilde q^+$ have no R charge, as follows from the action \p{6.7.1},  but they carry harmonic charge $1$. This implies that the six-point correlator  carries harmonic U(1) charges $2$ at points 1 to 4, and a total R charge 4 at points 0 and ${0'}$. Since the chiral $\theta_\alpha$ (at both the insertion and external points) are the only superspace coordinates with positive R charge $1/2$, we conclude that the correlator can
be written in the factorized form
\bea
\langle {L}{L}  O\tilde O O \tilde O \rangle
= \Theta(x,\q,u) \times f(x,u) + \mbox{$\bq$-terms}\, ,
\label{6pt}
\eea
where $\Theta$ is a particular nilpotent six-point superconformal covariant, homogeneous in $\q$ of degree 8, and thus carrying the whole R charge. The antichiral odd variables $\bq$ can only come from the external points, but we are ultimately interested only in the lowest components $\cO=O_{\q^+=\bq^+=0}$, so  we can ignore the $\bq$ terms in \p{6pt}. So, the essential information about the six-point correlator (\ref{6pt}) is contained in the function $f(x,u)$ without R charge.  Below we will show that this function is in fact harmonic independent, the harmonic $U(1)$ charge being carried by $\Theta$.

The structure of the nilpotent covariant $\Theta$ is determined by superconformal symmetry combined with the G-analytic nature of the four external points and the chiral nature of the two insertion points. In addition to Poincar\'e  supersymmetry (parameters $\epsilon^i_\alpha$, $\bar\ep_i^{\da}$) we need to consider
special conformal  supersymmetry (parameters $\eta^i_\alpha$, $\bar\eta_i^{\da}$).  In the chiral basis \p{chiba} we have
\begin{equation}\label{999}
    \delta x^{\alpha\dot\alpha}_L = -4i\theta^{i\alpha}\bar\epsilon^{\dot\alpha}_i
 -4i\theta^{i\alpha}x^{\beta\dot\alpha}_L
  \eta_{\beta i} \,, \qquad \delta \theta^{i\alpha} = \epsilon^{i\alpha} + x^{\alpha\dot\beta}_L
\bar\eta^i_{\dot\beta} + O(\q^2)\,,
\end{equation}
while in the G-analytic basis \p{ganba} we find
\begin{eqnarray}
 \delta x^{\alpha\dot\alpha}_A &=& -4iu^-_i(\epsilon^{i\alpha}
\bar\theta^{+\dot\alpha} +
\theta^{+\alpha}\bar\epsilon^{i\dot\alpha}) + 4i
(x^{\alpha\dot\beta}_A\bar\theta^{+\dot\alpha}\bar\eta^i_{\dot\beta}
-x^{\beta\dot\alpha}_A\theta^{+\alpha}\eta^i_\beta)u^-_i \, , \nonumber
\\
  \delta\theta^{+\alpha} &=& u^+_i\epsilon^{i\alpha} +
 x^{\alpha\dot\beta}_A\bar\eta^i_{\dot\beta}u^+_i
 + O(\q^2) \,, \qquad  \delta\bar\theta^{+\dot\alpha} = u^+_i\bar\epsilon^{i\dot\alpha}
- x^{\beta\dot\alpha}_A\eta^i_{\beta}u^{+}_i  + O(\bq^2) \, , \nonumber\\
 \delta u^+_i&=& 4i
(\theta^{+\alpha}\eta^j_\alpha
+\bar\eta^j_{\dot\alpha}\bar\theta^{+\dot\alpha})u^+_j u^-_i\,, \qquad
 \delta u^-_i =0 \;. \label{1000}
\end{eqnarray}
Now, the covariant $\Theta$ is homogeneous in $\q$ of degree 8. The inhomogeneous part of $\delta \q$,
\begin{equation}\label{linsusy}
    \delta \theta^{i\alpha} = \epsilon^{i\alpha} + x^{\alpha\dot\beta}_L
\bar\eta^i_{\dot\beta} \, , \qquad \delta\theta^{+\alpha} = u^+_i\epsilon^{i\alpha} +
 x^{\alpha\dot\beta}_A\bar\eta^i_{\dot\beta}u^+_i \, ,
\end{equation}
would lower this degree, unless we find combinations of $\q$'s which are invariant (to lowest order in $\q$, $\bq$) under $\cN=2$ superconformal supersymmetry. Such combinations are
\begin{equation}\label{xiv}
\xi_{r \dot \alpha} \, = \, \rho_{r\da}-\sigma_{r\da} \,, \qquad r=1,\ldots,4
\end{equation}
with
\begin{equation}\label{rhosig}
  \rho_{r\da} = (\theta_r^+ - \theta_0^i u_{ri}^+)^\a (x_{r0})_{\a\da}x^{-2}_{r0}\,, \qquad \sigma_{r\da} = (\theta_r^+ - \theta_{0'}^i u_{ri}^+)^\a (x_{r0'})_{\a\da}x^{-2}_{r0'}\,.
\end{equation}
Their total number is 8, and we wish to construct  the nilpotent covariant $\Theta$  of degree 8. We conclude that the leading term of $\Theta$ must involve all of  the variables $\xi_r$:
\begin{equation}\label{all8}
    \Theta = \xi^2_1\xi^2_2\xi^2_3\xi^2_4 \,.
\end{equation}

The aim of our two-loop calculation is to determine the factor $f(x,u)$ in the six-point correlator (\ref{6pt}). Since we are only interested in the lowest component of the four-point correlator $\vev{\cO\tilde \cO \cO \tilde \cO} =  \vev{O\tilde O O \tilde O}_{\theta^+=\bq^+=0}$, we can set all the external $\theta$s to zero, $\theta^+_r = 0$, $r=1,\ldots,4$. In this case $\Theta$ is rather simple \cite{Eden:2000mv}:
\begin{equation}\label{otherfr}
  \Theta|_{\theta^+= 0} = \theta_0^4\theta_{0'}^4\  \frac{(x_{0{0'}}^2)^2 \mathcal{R}}{\prod_{r=1}^4 x^2_{r0} x^2_{r0'}}\,,
\end{equation}
where
\begin{align}\label{R}
\mathcal{R} = (12)^2 (34)^2 x_{14}^2 x_{23}^2+(14)^2(23)^2 x_{12}^2x_{34}^2+(12)(23)(34)(41)\left[x_{13}^2x_{24}^2-x_{12}^2x_{34}^2-x_{14}^2x_{23}^2 \right]\,.
\end{align}

Finally, substituting everything into the double-insertion formula (\ref{ins}) and performing the trivial chiral integrations over $\theta_{0,{0'}}$, we obtain the two-loop correlator
\bea
 \vev{\cO\tilde \cO \cO \tilde \cO} =  \vev{O\tilde O O \tilde O}_{\theta^+=\bq^+=0} =
  {\cal R}
\int \frac{\rmd^4x_0 \rmd^4x_{0'}}{\prod_{r=1}^4 x^2_{r0} x^2_{r0'}} (x_{0{0'}}^2)^2\, f(x,u) \, .
\label{famp1}
\eea
Notice the characteristic presence of the polynomial prefactor ${\cal R}$. As shown in \cite{Eden:2000bk,Arutyunov:2003ae,Arutyunov:2003ad}, this factorization of the loop corrections is a universal feature, called ``partial non-renormalization".

\subsubsection{Feynman graphs. Harmonic analyticity}\label{Fgha}

Now, the practical question is how to compute $f(x,u)$ from the corresponding set of two-loop Feynman diagrams. It turns out that instead of setting $\theta^+=0$, as required in the final expression (\ref{famp1}), it is much more convenient to do the computations with $\theta_{0,{0'}}=0$. The knowledge of the complete $\Theta$ (\ref{all8}) allows us to easily switch from one of these forms to the other. The new form of $\Theta$ is even simpler, yielding
\begin{equation}\label{chk}
 \langle {L}{L}  O\tilde O O \tilde O \rangle_{\theta_{0,{0'}}=\bq^+=0} = (x_{0{0'}}^2)^4\prod_{r=1}^4
  \frac{(\theta^+_r)^2}{x^2_{r0} x^2_{r0'}}\, f(x,u) \, .
\end{equation}
Then it is clear that in working out the expressions for the various Feynman graphs we can concentrate only on the terms with the maximal number of external $\theta^+$.
In particular, at order $g^4$ this choice removes all graphs which contain non-Abelian interaction vertices. For example, the non-Abelian gluon subgraph in Fig.~\ref{fig5}(c)  vanishes because it has two chiral ends at the insertion points 0 and ${0'}$ and one G-analytic end (the gluon without insertion); after setting $\theta_{0,{0'}}=0$ we are left with too few chiral $\theta^+$s at the G-analytic gluon end to supply the required R charge 2. Similarly, the   block in Fig.~\ref{fig5}(d) has three chiral ends (in fact, only two, points ${0'}$ and $0''$ should be identified) and two G-analytic ends; once again, the G-analytic $\theta^+$s cannot provide the required R charge 3. The same applies to the block  in Fig.~\ref{fig5}(e).

\begin{figure}[h!]
\psfrag{2}[cc][cc]{$\scriptstyle 2$}
\psfrag{1}[cc][cc]{$\scriptstyle 1$}
\psfrag{0}[cc][cc]{$\scriptstyle 0$}
\psfrag{0'}[cc][cc]{$\scriptstyle 0'$}
\psfrag{0''}[cc][cc]{$\scriptstyle 0''$}
\psfrag{0'''}[cc][cc]{$\scriptstyle 0'''$}
\psfrag{(a)}[cc][cc]{(a)}
\psfrag{(b)}[cc][cc]{(b)}
\psfrag{(c)}[cc][cc]{(c)}
\psfrag{(d)}[cc][cc]{(d)}
\psfrag{(e)}[cc][cc]{(e)}
%
\centerline{ \includegraphics[width=\textwidth]{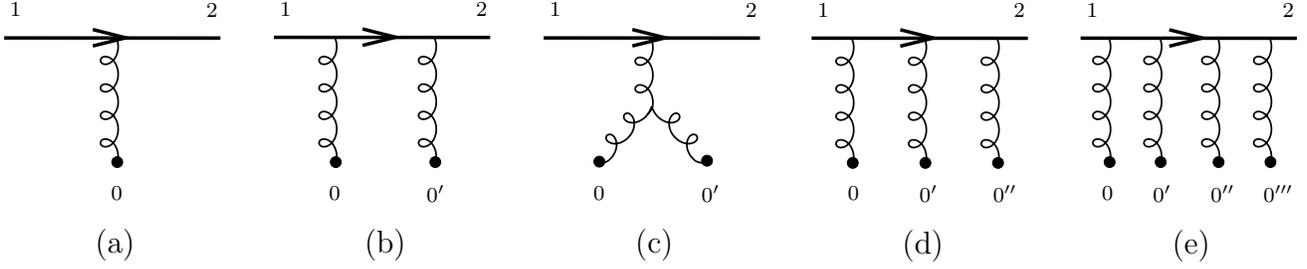} }
 \caption{\small  Building blocks of the Feynman graphs. }\label{fig5}
 \end{figure}

As a result of all these simplifications our task is reduced to
listing all {\it tree level} Feynman graphs made out of the two
building blocks T and TT in Figs.~\ref{fig5}(a) and (b).  They are
calculated with the Feynman rules from Sect.~\ref{FR}. Although
these blocks contain gauge/matter interaction vertices \p{vertex}
and hence space-time integrals, the latter are easily done using
the identities
\bear \square_1\int {d^4x_3\over x^2_{13}
x^2_{23}x^2_{03}} &=& {4i\pi^2 \over x^2_{12} x^2_{10}}\;,  \qquad
\partial^{[\mu}_1\partial^{\nu]}_2\int {d^4x_3\over x^2_{13} x^2_{23}x^2_{03}} =
-4i\pi^2 {x^{[\mu}_{10}x^{\nu]}_{20}\over x^2_{12}
x^2_{10}x^2_{20}}\,,
\label{reduceint}
\ear
producing very simple {\it rational} space-time functions \cite{Howe:2000hz,Eden:2000mv}:
\begin{equation}\label{T}
  \langle \tilde q^+_a(1) W_b(0) q^+_c(2) \rangle = - \frac{2 i g^2 f_{abc}}{(2\pi)^4\;  x_{12}^2}\; \Big[(21^-)\rho_1^2+(12^-)\rho_2^2 -2(\rho_1\rho_2)\Big]
\end{equation}
\begin{equation}\label{TT}
   \langle \tilde q^+_a(1) W_b(0) W_d({0'})  q^+_e(2) \rangle \, = \, -\frac{4
   g^4 f_{abc}f_{cde}}{(2\pi)^6\; x_{12}^2} \, (1^-2^-)\rho_1^2\, \sigma_2^2\,,
\end{equation}
where $\rho$ and $\sigma$ were defined in \p{rhosig}.

Notice the characteristic presence of negative-charged harmonics in both expressions (\ref{T}) and (\ref{TT}). This has to do with the important issue of harmonic analyticity \cite{Eden:2000qp}. In an interacting theory the hypermultiplet satisfies its equation of motion \p{EMo} with a covariant harmonic derivative, $D^{++} q^+ + ig [ V^{++},q^+]=0$.  The gauge-invariant composite operators ${O}_k={\rm Tr} (q^+)^k$ satisfy the same equation with a flat harmonic derivative, $D^{++}{O}_k=0$. As explained above, the harmonic derivative is the raising operator of $SU(2)$. So, ${O}_k$ corresponds to the highest-weight state of an SU(2) irrep of weight $k$ (a $(k+1)$-plet).

In practice, this means that the $n-$point correlator is annihilated by the harmonic derivative $D^{++}$ at each point,
\begin{equation}\label{haco}
    D^{++}_r \vev{{O}_k\ldots{O}_k} =0\,, \qquad r=1,\ldots, n\,.
\end{equation}
Since $D^{++} u^+=0$ and $D^{++} u^-=u^+$, this implies that the correlator is a polynomial in $u^+_r$, $r=1,\ldots,n$,  homogeneous of degree $k$ at each point, and no dependence on $u^-$ is allowed.

Clearly, the expressions for the building blocks T (\ref{T}) and
TT (\ref{TT}) are
not harmonic analytic because of the presence of $u_1^-$ and $u_2^-$.
This, however,
is not a problem: The various building blocks or even complete Feynman
graphs are
not expected to be harmonic analytic, much like they are not conformal and gauge
invariants.
It is only the sum of all graphs that has these properties. Indeed, it can be shown that by summing up all graphs made from the T and TT blocks, all negative-charged harmonics drop out. To see this one uses the harmonic cyclic identity, e.g.,
\begin{equation}
(12)(1^-2^-)   -   (12^-)(1^-2)    = u^{+i}_1 u^{-j}_1 u^{+k}_2 u^{-l}_2 (\ep_{ik} \ep_{jl} - \ep_{il} \ep_{jk}) =  (11^-)(22^-)   =   1\,,  \label{idEx}
\end{equation}
as a consequence of the defining property $u^{+i} u^-_i = 1$ \p{defharvar}.
In practice, the use of the cyclic identity is cumbersome when there are $u$'s
from too many different points.
But we can do better, by completely sidestepping this issue.

We can profit from the expected harmonic analyticity
of the final result to greatly simplify our graph calculations. Let us come back to the correlator of four operators ${O} \equiv {O}_2$ with two Lagrangian insertions, calculated at $\q_0=\q_{0'}=0$, see \p{chk}. The Lagrangian has no  harmonic $U(1)$ charge, hence $\vev{{L}{L} {O}\tilde{O}{O}\tilde{O}}$ should have charges $+2$ at each external point.  From \p{chk} we see that the nilpotent factor already carries the necessary charges, thus making the function $f(x,u)$ chargeless. Harmonic analyticity then implies that this function is {\it harmonic independent}. This allows us to compute the correlator  \p{chk} with all four harmonic variables identified,
\begin{equation}\label{idha}
    u^\pm_1=u^\pm_2=u^\pm_3=u^\pm_4\,.
\end{equation}
This simple trick eliminates a number of irrelevant Feynman graphs, namely, all graphs with at least one free hypermultiplet line, since the hypermultiplet propagator \p{qproat0} vanishes if $u^+_1=u^+_2$. Among them we find the graphs with the  blocks from Fig.~\ref{fig5}(c)-(e), for which we already gave a different reason why they do not contribute. In addition, the identification of harmonics eliminates the graphs with TT blocks.   This leaves  only the three graphs shown in Fig.~\ref{survivors} (plus point permutations):

\begin{figure}[h!]
\psfrag{2}[cc][cc]{$\scriptstyle 2$}
\psfrag{1}[cc][cc]{$\scriptstyle 1$}
\psfrag{0}[cc][cc]{$\scriptstyle 0$}
\psfrag{0'}[cc][cc]{$\scriptstyle 0'$}
\psfrag{0''}[cc][cc]{$\scriptstyle 0''$}
\psfrag{0'''}[cc][cc]{$\scriptstyle 0'''$}
\psfrag{(a)}[cc][cc]{(a)}
\psfrag{(b)}[cc][cc]{(b)}
\psfrag{(c)}[cc][cc]{(c)}
\psfrag{(d)}[cc][cc]{(d)}
\psfrag{(e)}[cc][cc]{(e)}
%
\centerline{ \includegraphics[width=0.9\textwidth]{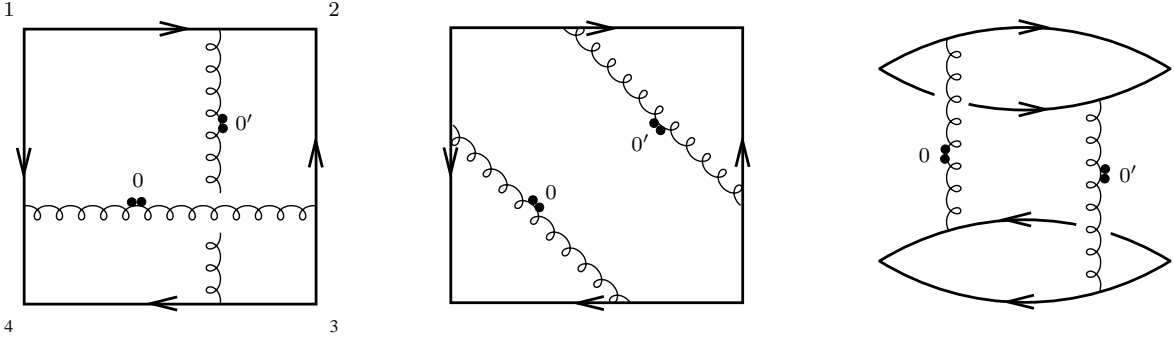} }
 \caption{\small  \label{survivors} Graphs surviving the identification of harmonics.}
 \end{figure}

\noindent Notice that despite the appearance, the third graph is planar. This has to do with the fact that unlike the Green's functions of elementary fields, which are planar on a disk, the correlators (Green's functions) of composite operators have the topology of a sphere. This peculiar property of the correlators was mentioned in Sect.~\ref{Dure}.

Each of the graphs in Fig.~\ref{survivors} is a product of T blocks \p{T} evaluated at $\q_0=\q_{0'}=0$. Thus, the calculation of the harmonic-independent function $f(x)$ in \p{chk} is reduced to elementary algebra. Then $f(x)$ is substituted in \p{famp1}, leading to the final result \p{Born2} from \cite{Eden:2000mv}.

This procedure is a very efficient tool for calculating not only four-point, but  also $n$-point correlators of half-BPS operators made of hypermultiplets. In the next subsection we show how to adapt the procedure to five points.

\subsection{The insertion procedure: Five points}\label{ip5p}

Here we evaluate the correlator
\begin{equation}
G_{5+2} \, = \, \langle { {L}(0) \; {L}({0'})\; \cal O}(1) \; \tilde {O}(2) \;
{O}(3) \; \tilde {O}(4) \; \hat{O}(5)
\rangle\,,
\end{equation}
with ${O}   = \mathrm{Tr}(q^+)^2,    \tilde {O}   =
\mathrm{Tr}(\tilde q^+)^2,   \hat{O}   =   2 \, \mathrm{Tr}(\tilde
q^+ q^+),    {L} = 1/(2 g^2)\, {\rm Tr}(W)^2$, in order to study
the two-loop corrections to the correlator
\begin{equation}
{G}_{5} \, = \, \langle {O}(1) \; \tilde {O}(2) \;
{O}(3) \; \tilde {O}(4) \; \hat{O}(5) \rangle \, ,
\end{equation}
discussed in Sect.~\ref{52loop}.
We will use the technology developed in \cite{Eden:2000mv} and reviewed in Appendix~\ref{ip4p} for the
four-point case $G_{4} \, = \, \langle {O}(1) \; \tilde {O}(2) \;
{O}(3) \; \tilde {O}(4) \rangle$.

As before, we compute $G_{5+2}$ in $D=4$. It remains finite in the light-cone limit
\begin{equation}
x_{12}^2 \, = \, x_{23}^2 \, = \, x_{34}^2 \, = \, x_{45}^2 \, = \, x_{51}^2 \,
\rightarrow 0 \, .
\end{equation}
The divergences of $G_5$ in this limit arise from the integration over
the insertion points $x_0 , \, x_{0'}$ which we will eventually regularize by the
IR prescription of changing only the dimension of the integration
measure at these points.

Repeating the superconformal argument of Appendix~\ref{snsc}, we can claim that the leading term in the $\q$ expansion of the correlator $G_{5+2}$ will factorize into a  nilpotent covariant $\Theta \sim \q^8$ and some function $f(x,u)$ of the bosonic variables, see \p{6pt}. Once again, $\Theta$ will depend
only on the invariant variables $\xi_r$ \p{xiv} (with $r=1,\ldots,5$). This time, however, we have 10 such variables, while the  degree of homogeneity of $\Theta$ is still 8. Unlike the four-point case, where the nilpotent structure \p{all8} was unique, now we can have two different choices:
\begin{eqnarray}
A_5 & = & \xi_1^2 \, \xi_2^2 \, \xi_3^2 \, \xi_4^2 \, f_5(x,u) \, \nonumber \\
B_{45} & = & \xi_1^2 \, \xi_2^2 \, \xi_3^2 \, \xi_{4 \dot \alpha} \, \xi_{5 \dot
\b} \, f_{45}^{\dot \alpha \dot \b}(x,u)\,,
\end{eqnarray}
and their point permutations. A graph calculation is needed to fix the
coefficient functions $f(x,u)$. The knowledge that only $A_r, \, B_{rs}$ can occur is
very useful, though, if combined with the property of harmonic analyticity,
namely the fact that the full gauge invariant correlator only depends
on $u^+_r$ and that it does so in a polynomial fashion (see Appendix~\ref{Fgha}).

Let us consider the covariants $A_r, \, B_{rs}$ at $\theta_0 = \theta_{0'} = 0$,
in other words,  only the terms involving $\theta_r^+$ from the outer
points. Now, $\theta_{r}^+ $ carries $U(1)$ charge 1 at point $r$, and
so does $u^+_{r}$. Hence the spinor part of
the  covariant $A_5$ has charge 2 at points 1,2,3,4, whereas it is
chargeless at point 5.
Since all five outer operators ${O}, \tilde {O}, \hat{O}$
carry charge 2, the coefficient function $f_5(x,u)$ has to have charge 2 at
point 5 and zero at all other points. To be harmonic analytic it must be an $SU(2)$ invariant polynomial
in the $u^+_r$ with the correct charges. The only such invariant is
trivially zero, $
(55) \, = \, \epsilon^{ij} u^+_{5j} u^+_{5i} \, = \, 0 $.
This argument rules out all $A_r$ covariants.

Further, in $B_{45}$ the odd variables $\xi$ carry charge 2 at points 1,2,3 and charge 1 at
points 4,5. Due to harmonic analyticity, the harmonic dependence of the bosonic factor can only be of the form $f_{45}^{\dot \alpha \dot \b}(x,u)=(45)f_{45}^{\dot \alpha \dot \b}(x)$. Thus, the sum of the contributions of the graphs to this covariant will be
\begin{equation}\label{b45}
B_{45} \, = \, \xi_1^2 \, \xi_2^2 \, \xi_3^2 \, \xi_{4 \dot \alpha} \, \xi_{5 \dot
\b} \,  (45) \, f_{45}^{\dot \alpha \dot \b}(x)  \, .
\end{equation}
Individual graphs do contain non-analytic terms, but we need not go through all the details of how they cancel out. Instead, we can apply the powerful trick of identifying the harmonic variables,  as we did in the four-point case in Appendix~\ref{Fgha}.

For each covariant $B_{rs}$ (with $r,s=1,\ldots,5$)
we know that the harmonic dependence of the coefficient function will eventually be given by just $(rs)$. This result clearly does not change if all harmonics, except for
$u^\pm_r$, are put equal to $u^\pm_s$. As a convention, if $r < s$ we will keep
$u_{ri}^\pm$ aside and identify all other harmonics with $u_{si}^\pm$. We
will obtain the correct result if this is done consistently for any
contribution to the spinor structure pertaining to the given covariant \p{b45}.

This manoeuvre drastically
simplifies the use of the harmonic cyclic identity. For example equation (\ref{idEx})
reduces to $0 \, - (-1) \, = \, 1$ if both $u^\pm_{1 i}, \, u^\pm_{2 i}$
are sent to $u^\pm_{5 i}$. But there are more far-reaching consequences for the supergraphs at $\theta_0 = \theta_{0'} = 0$.
\begin{itemize}
\item Any diagram with more than one free line (i.e. a hypermultiplet propagator
between two outer points) is put to zero. Spinor structures relating to the
$A_r$ covariants may be discarded immediately. For the $B_{rs}$ type
contributions the suggested identification of the harmonics will send at
least one of the numerators of the free lines to $(ss) = 0$.
\item As a consequence, we only need to take into account graphs built out of four
T blocks and one free hypermultiplet line.
\item The light-cone limit singles out diagrams
in which the gluon lines connect to the pentagon frame 123451 of matter lines,
because the T blocks with outer ends $r,s$ have an explicit propagator factor $1/x_{rs}^2$. This eliminates graphs with disconnected matter frames, like the third graph in Fig.~\ref{survivors}, since they lack the required light-cone singularity. For the same reason, graphs with a connected ``zigzag" frame like in Fig.~\ref{ttree}(b) are not allowed.
\end{itemize}

All of these simplifications leave us with a very small number of graphs shown in Fig.~\ref{try} (notice that the middle graph is planar, like the two others).
\begin{figure}[h!]
\psfrag{2}[cc][cc]{$\scriptstyle 2$}
\psfrag{1}[cc][cc]{$\scriptstyle 1$}
\psfrag{0}[cc][cc]{$\scriptstyle 0$}
\psfrag{0'}[cc][cc]{$\scriptstyle 0'$}
\psfrag{0''}[cc][cc]{$\scriptstyle 0''$}
\psfrag{0'''}[cc][cc]{$\scriptstyle 0'''$}
\psfrag{(a)}[cc][cc]{(a)}
\psfrag{(b)}[cc][cc]{(b)}
\psfrag{(c)}[cc][cc]{(c)}
\psfrag{(d)}[cc][cc]{(d)}
\psfrag{(e)}[cc][cc]{(e)}
%
\centerline{ \includegraphics[width=\textwidth]{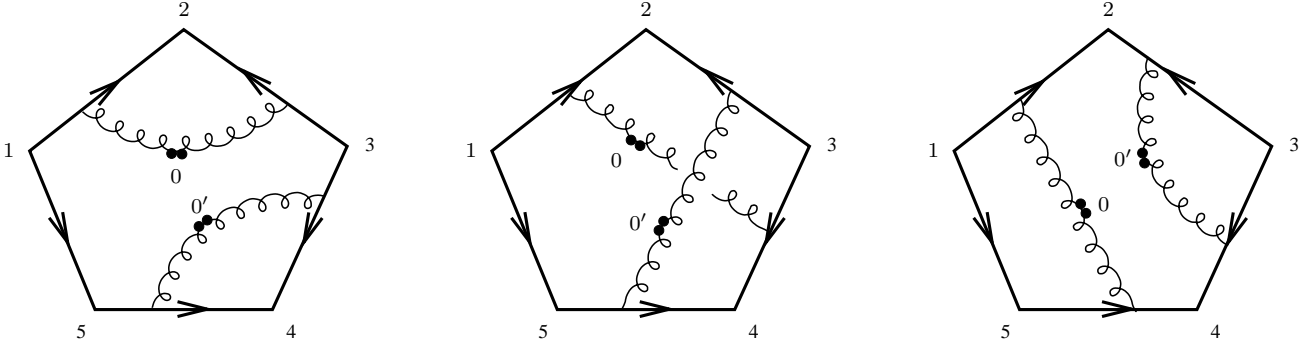} }
 \caption{\small  \label{try} Graphs contributing to $G_{5+2}$. The permutations of the insertion points $0$, $0'$ and the cyclic permutations of the external points should be added.}
 \end{figure}

\noindent The resulting algorithm for evaluating these graphs is as follows:
\begin{itemize}
\item For any of the diagrams in this set multiply out the terms of the four T
blocks and classify them according to the various $\theta_r^+$ structures. Discard terms with four $(\q^+_r)^2$ (A type)
but keep those related to the B type covariants. Sum over all graphs.
\item In each term identify the harmonics according to the spinor structure;
by way of example for $(\theta_1^+)^2 (\theta_2^+)^2 (\theta_3^+)^2
\theta_{4 \alpha}^+ \theta_{5 \b}^+$ we put
$u^\pm_{1 i} = u^\pm_{2 i} = u^\pm_{3 i} = u^\pm_{5 i}$.
The contributions of some diagrams may vanish in doing so, in other cases the
harmonics will reduce to the simple factor (45).
\item Reconstruct the full covariants. The result of the procedure is the
entire leading term of the correlator $G_{5+2}$ in which the harmonics
are not identified any longer. We can now switch to the ``opposite end'' of
the expression by putting all $\theta_r^+ = 0$, while
restoring the spinors from the insertion points. We observe that this step produces
$(12)(23)(34)(45)(51) \, \theta_0^4 \, \theta_{0'}^4 \, g_0$, i.e. in the light-cone
limit no other $SU(2)$ channel is present.
\end{itemize}
The reconstruction is in fact elementary. From the definitions \p{xiv}, \p{rhosig} we find
\begin{equation}
\theta_r^{+ \alpha} \, = \, \frac{x_{r0}^2 \, x_{r{0'}}^2}{x_{0{0'}}^2}
\left( \frac{x_{r0}^{\alpha \dot \alpha}}
{x_{r0}^2} - \frac{x_{r{0'}}^{\alpha \dot \alpha}}{x_{r{0'}}^2} \right) \,
\xi_{r \dot \alpha}|_{\theta_0 = \theta_{0'} = 0} \, .
\end{equation}
By this formula we can unambiguously upgrade every $\theta_r^+$ from
the graph calculation to the invariant combination $\xi_r$.
Next we note that
\begin{equation}
\frac{x_{r0}^2 \, x_{r{0'}}^2}{x_{0{0'}}^2}
\left( \frac{x_{r0}^{\alpha \dot \alpha}}
{x_{r0}^2} - \frac{x_{r{0'}}^{\alpha \dot \alpha}}{x_{r{0'}}^2} \right) \,
\xi_{r \dot \alpha}|_{\theta_r^+ = 0} \, = \, x_{r0}^{\alpha \dot \alpha} \,
\lambda_{{0'} \dot \alpha}^{r+} \, - \, x_{r{0'}}^{\alpha \dot \alpha} \,
\lambda_{0 \dot \alpha}^{r+}
\end{equation}
where
\begin{equation}
\lambda_{t \dot \alpha}^{r+} \, = \, \frac{x_{0{0'} \, \dot \alpha \alpha}}
{x_{0{0'}}^2} \, \theta_t^{\alpha i} u^+_{r i} \, , \qquad t \in \{0,{0'}\} \, .
\end{equation}
In order to complete the task we must collect the $u$-projected $\lambda$
variables into $\theta_0^4 \, \theta_{0'}^4$ and a harmonic factor. To this end we use the identity
\begin{eqnarray}
\lambda_{0 \dot \alpha \phantom{\dot \beta}}^{1+}
\lambda_{0 \dot \beta \phantom{\dot \beta}}^{2+}
\lambda_{0 \dot \gamma \phantom{\dot \beta}}^{3+}
\lambda_{0 \dot \delta \phantom{\dot \beta}}^{4+}
& = & \frac{1}{(12)^2} \, \lambda_{0 \dot \alpha \phantom{\dot \beta}}^{1+}
\lambda_{0 \dot \beta \phantom{\dot \beta}}^{2+} \left( (13)
\lambda_{0 \dot \gamma \phantom{\dot \beta}}^{2+} - (23)
\lambda_{0 \dot \gamma \phantom{\dot \beta}}^{1+} \right)
\left( (14)
\lambda_{0 \dot \delta \phantom{\dot \beta}}^{2+} - (24)
\lambda_{0 \dot \delta \phantom{\dot \beta}}^{1+} \right) \\
& = & \frac{1}{4 \, (x_{0{0'}}^2)^2} \left( \epsilon_{\dot \alpha \dot \gamma}
\, \epsilon_{\dot \beta \dot \delta} \, (14)(23) -
\epsilon_{\dot \alpha \dot \delta} \, \epsilon_{\dot \beta \dot \gamma} \,
(13)(24) \right) \, \theta_0^4\,,
\nonumber
\end{eqnarray}
and its special cases where some points coincide.
Beyond the factor (rs) in the coefficient function of the $B_{rs}$ covariants,
the conversion to $\theta_0^4 \, \theta_{0'}^4$ produces four further harmonic
factors (ij) which now carry the remaining $U(1)$ charges. The
harmonic dependence remains manifestly analytic. Out of the $u_r^+$ from
the five outer
points one can construct six independent polynomials carrying charge 2 at every
point. We stress that it is not obvious that the sum of graphs produces only
one channel in the light-cone limit, namely $(12)(23)(34)(45)(51)$.

In summary, we have explained how the evaluation of the correlator
$G_{n+2}$ is reduced to algebraic manipulations by the insertion
procedure combined with superconformal symmetry and harmonic analyticity.
As in the four-point case, no integral needs to be done once the T block is known.
Nevertheless, the amount of algebra is fairly large, so that we have resorted
to a \emph{Mathematica} script. It remains to restore the integrations over the isnertion points $x_0, x_{0'}$ with the IR-modified measure. The scalar (parity-even) part of the result for $G_5$ has the concise form displayed in Eq.~\p{fin}.

A special comment is due here on the pseudo-scalar (parity-odd) part of the correlator. Our calculation of $G_{5+2}$ does indeed produce such a (rather complicated) part. But this does not mean that $G_5$ will have a parity-odd part. It must drop out after the integration over the insertion points $x_0$, $x_{0'}$. The explanation is given in Appendix~\ref{pary}.

\subsection{Parity properties of the scalars in the $\cN=2$ theory}\label{pary}

Here, following Ref.~\cite{Ogievetsky:1988av}, we argue that there exists a parity assignment for the fields of the $\cN=2$ vector and hypermultiplets, such that the hypermultiplet scalars are true scalars (not pseudo). With this assignment, all our operators made of hypermultiplets are scalars, and their correlators should not contain a parity-odd part.

The components of the two multiplets
are contained in the G-analytic superfields $q^+$, Eq.~\p{onshq'}, and $V^{++}$, Eq.~\p{6.10.1}. The parity assignments of Ref.~\cite{Ogievetsky:1988av} for the superspace coordinates are
\begin{equation}\label{par}
   P: \qquad x'_0 = x_0\,, \quad {\vec{x}}' = - \vec{x}\,, \quad (\q^+_\a)' = \bq^{+\da}\,, \quad (\bq^+_{\da})' = - \q^{+\a}\,, \quad (u^\pm_i)' = u^\pm_i\,,
\end{equation}
while the superfields remain {\it inert},
\begin{equation}\label{parsup}
    {q^+}'(x',\q',u') = q^+(x,\q,u)\,, \qquad {V^{++}}'(x',\q',u') = V^{++}(x,\q,u)\,.
\end{equation}
In terms of the bosonic physical fields, these assignments imply that the hypermultiplet scalars $\phi^i$ are {\it true scalars}, the gluon $A_\mu$ is a polar vector, while the complex vector multiplet scalar $\varphi$ is a mixture of a true and a pseudo-scalar. In addition, the fermion fields transform in an unusual way: the hyperinos transform into each other (up to signs), $\psi  \leftrightarrow \bar\kappa$, and the gluinos $\lambda^i    \leftrightarrow \bar\lambda^i$. The latter relation means that, e.g., $\lambda^1    \leftrightarrow \bar\lambda^1 = \bar\lambda_2 = (\lambda^2)^*$, which again differs from the traditional assignment $\lambda^i    \leftrightarrow (\lambda^i)^*$.

These rules can be tested for consistency by inspecting the Yukawa couplings in the $\cN=4$ Lagrangian. The gauge/matter coupling reads
\begin{equation}\label{GM}
    \int du d^4x d^4\q^+ \ {\rm Tr} (V^{++}[\tilde q^+, q^+])\ \Rightarrow\ {\rm Tr}[\phi^i(\{\kappa, \lambda_i\} + \{\bar\psi, \bar\lambda_i\})] + {\rm c.c.}
\end{equation}
We see that the above assignments allow $\phi^i$ to stay inert under parity. At the same time, the
Yukawa coupling from the gauge sector is
\begin{equation}\label{gyuk}
   \int d^4x d^4\q \ {\rm Tr} (W^2) \ \Rightarrow\  \int d^4x {\rm Tr}\left(\bar\varphi\{\lambda^i, \lambda_i\} + \varphi\{\bar\lambda_i, \bar\lambda^i\}\right)\,.
\end{equation}
Here the combination of fermions accompanying each boson is chiral, therefore $\varphi    \leftrightarrow \bar\varphi$.

The above parity assignments  mean that the hypermultiplet composite operators of the type $\cO={\rm Tr}(q^+)^k|_{\q=0} = {\rm Tr}(\phi^{i_1}(x) \ldots \phi^{i_k}(x)) u^+_{i_1} \ldots u^+_{i_k}$ are all true scalars. Thus, the correlators $\vev{\cO\tilde \cO \cO \tilde \cO}$ that we are considering cannot have a parity-odd part.

The reason why we see such a part in the correlators $G_{n+1}$ and $G_{n+2}$ was explained after Eq.~\p{pseli}. It is due to the insertion of the complex (chiral) form of the SYM Lagrangian. In it we find pseudo-scalar terms, for example $iF\tilde F$, which are responsible for the parity-odd part in the correlators with insertions. But at the final stage of the calculation, the integration over the insertion point will eliminate all such terms, which are total derivatives. Indeed, we have already encountered this phenomenon in Sect.~\ref{n1l}. The correlator with one insertion $G_{n+1}$ has the form \p{pseli}, where we clearly see a pseudo-scalar in the last line. However, it drops out after the integration over the insertion point.

\section{Four-point correlators of operators of weight $k$}\label{multiap}

Let us consider the four-point correlator of protected half-BPS operators of weight $k$. At the lowest level of the $\q$ expansion they are built from $k$
elementary scalar fields, ${\rm Tr}(\phi^k)$. Such correlators are discussed in detail in \cite{Arutyunov:2003ae,Arutyunov:2003ad}, using the method of \cite{Eden:2000mv}. We consider the following $\cN=2$ hypermultiplet projection
\begin{equation}\label{trlo}
   G_{4;k}= \vev{\cO^k(1) \tO^k(2) \cO^k(3) \tO^k(4)}
\end{equation}
where in this appendix $\cO^k = 2^{k/2} {\rm Tr}(q^+)^k$.

The (connected, planar) tree-level correlator is given by (see Fig.~\ref{tree})
\begin{equation}\label{tre}
    G^{(0)}_{4;k} = k^4 N_c^{2k-2}\sum_{m=1}^{k-1} X^m Y^{k-m}\,,
\end{equation}
where the notation was introduced for the harmonic and space-time propagator factors
\begin{align}
X= \frac{(12)(34)}{(2 \pi)^4 x_{12}^2 x_{34}^2}\,,
\quad Y= \frac{(14)(32)}{(2 \pi)^4 x_{14}^2 x_{23}^2}\,.
\end{align}

\begin{figure}[h]
\psfrag{m}[cc][cc]{$\scriptstyle m$}
\psfrag{k-m}[cc][cc]{$\scriptstyle k-m$}
\psfrag{1}[cc][cc]{$\scriptstyle 1$}
\psfrag{2}[cc][cc]{$\scriptstyle 2$}
\psfrag{3}[cc][cc]{$\scriptstyle 3$}
\psfrag{4}[cc][cc]{$\scriptstyle 4$}
%
\centerline{ \includegraphics[height=50mm]{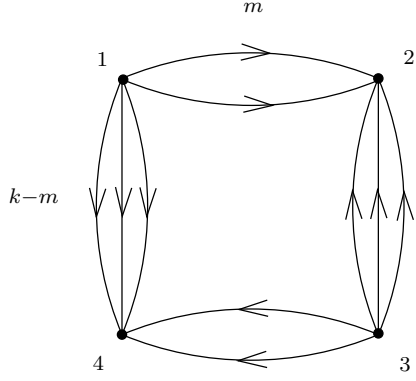} }
 \caption{\small  Tree graphs for operators of weight $k$. \label{tree}  }
 \end{figure}

The loop corrections have the general {\it factorized} form
\begin{align}\label{gen}
G_{4;k}^{(\rm loops)} = {\cal R}' \sum_{m=0}^{k-2}
\mathcal{F}^k_{m}(x) X^m Y^{k-m-2}
\end{align}
with the polynomial prefactor
\begin{align}
{\cal R}' =  sX^2 + (t-s-1) XY + Y^2
\end{align}
involving  the conformal cross-ratios
\begin{equation}\label{cccr}
    s=\frac{x_{12}^2x_{34}^2}{x_{14}^2x_{23}^2}\,,\quad t=\frac{x_{13}^2x_{24}^2}{x_{14}^2x_{23}^2}\,.
\end{equation}
The coefficient functions $\mathcal{F}(x)$ will be specified below.

The result of \cite{Arutyunov:2003ae} for the one-loop correction in the planar limit is
\begin{equation}\label{onel}
    G_{4;k}^{(1)}  = k^4 N_c^{2k-2} {\cal R}'\  2 a \, x^2_{14} x^2_{23}\, g(1234) \sum_{m=1}^{k-1} X^{m-1} Y^{k-m-1}\,,
\end{equation}
where the one-loop box integral $g(1234)$ is defined in \p{inth}. We see that in this case the general amplitude \p{gen} becomes completely degenerate, with all $\mathcal{F}^k_{m}(x) \propto x^2_{14} x^2_{23}\, g(1234)$.
Going to the light-cone limit $x^2_{12}=x^2_{23}=x^2_{34}=x^2_{41} \to 0$, we find that $s$ remains finite while $t \to \infty$, hence
\begin{align}\label{Rlc}
{\cal R}' \to  t XY\,.
\end{align}
As a consequence, Eq.~\p{onel} simplifies to
\begin{equation}\label{onel'}
    G^{(1)}_{4;k}  =  k^4 N_c^{2k-2} \ 2 a \, x^2_{13} x^2_{24}\, g(1234) \sum_{m=1}^{k-1} X^m Y^{k-m} + \mbox{subleading terms}\,.
\end{equation}
Dividing this expression by the tree-level correlator \p{tre}, we obtain the same ratio as in the case $k=2$ at one loop, see \p{ratio}. Thus, the ratio does not depend on the value of $k$, up to one loop.

The same pattern is found at two loops. According to \cite{Arutyunov:2003ad}, the two-loop coefficient functions in \p{gen} are
\begin{align}\notag
\mathcal{F}^k_{m} = \frac{g^4}{(8 \pi^2)^2} \big\{& \left[ C_m^d x^2_{13} x^2_{24} + (2s C_m^b-C_m^d)x^2_{14} x^2_{23}+\left({2} C_m^{b}/s-C_m^d\right)x^2_{12} x^2_{34}\right] [g(1234)]^2
\\ \notag
&  +(C_m^c-C_m^{d})2[x^2_{13}h(123;134) + x^2_{24}h(124;234)] \\ \notag
&  +(C_m^d-C_m^{a_1})2[x^2_{14}h(124;134) + x^2_{23}h(123;234)] \\ &   +(C_m^d-C_m^{a_2})2[x^2_{12}h(123;124) + x^2_{34}h(134;234)] \big\}\,,  \label{planar}
\end{align}
with the two-loop integral $h$ defined in  \p{inth}  and with color factors $C$ given in  \cite{Arutyunov:2003ad}. Going to the light-cone limit, only the first term from the first line and the term from the second line survive. Further, in the planar limit the remaining color factors simplify to (see \cite{Arutyunov:2003ad})
\begin{align}
C_m^d = k^4 N_c^{2k}\,,\qquad C_m^c = 2k^4 N_c^{2k}\,.
\end{align}
As a result, we find that on the light cone all the relevant two-loop coefficients become equal,
\begin{align} \label{planar'}
& \mathcal{F}^k_{m}(x) = a^2 \, k^4 N_c^{2k-2}  \left\{  (x^2_{13} x^2_{24})^2 [g(1234)]^2 + 2x^2_{13} x^2_{24}[x^2_{13}h(123;134) + x^2_{24}h(124;234)] \right\} \,.
\end{align}
Substituting this result into \p{gen} and taking account of \p{Rlc}, we obtain the two-loop correction
\begin{align} \notag
 G^{(2)}_{4;k}  = &
 \big\{  (x^2_{13} x^2_{24})^2 [g(1234)]^2 + 2x^2_{13} x^2_{24}[x^2_{13}h(123;134) + x^2_{24}h(124;234)] \big\}
 \\ 
 &   \times a^2 \, k^4 N_c^{2k-2} \sum_{m=1}^{k-1} X^m Y^{k-m} + \text{subleading terms}\,.
\end{align}
Finally, dividing by the tree-level correlator \p{tre}, we obtain the same ratio as in the case $k=2$ at two loops, see \p{ratio}.  This confirms that the ratio does not depend on the value of $k$ up to two loops.

In conclusion, we can claim that the duality relation
\begin{equation}\label{relcak}
   \lim_{x^2_{i,i+1}\to 0} G_{4;k}/G_{4;k}^{(0)}  = \lr{\mathcal{A}_{4;k}/\mathcal{A}_{4;k}^{(0)}}^2 + O(a^3)
\end{equation}
holds for any weight $k$ of the half-BPS operators.

\subsubsection{Relation to Wilson loops}\label{app:W}

From our analysis it follows that the correlator in the planar limit has the following universal form on
the light-cone
\begin{align}
\vev{  q^{k_1} (1) \tilde q^{k_2}(2) q^{k_3}(3) \tilde q^{k_{4}}(4)} =\vev{  q^{k_1} (1) \tilde q^{k_2}(2) q^{k_3}(3) \tilde q^{k_{4}}(4)}^{(0)} \left[ W (x_1 , x_2 , x_3 , x_{4} )\right]^2\,,
\end{align}
where $W (x_i)$ is a light-like Wilson loop in the fundamental representation of $SU (N_c)$ evaluated along the contour $[x_1 , x_2 ] \cup [x_2 , x_3 ] \cup [x_3 , x_{4} ] \cup [x_{4} , x_1 ]$,
\begin{align}
 W(x_i)=\frac1{N_c}\vev{0|\Tr {\rm P}\exp \lr{ig\oint_\square dx\cdot A(x) }|0}\,.
\end{align}
\begin{figure}[h!]%
\psfrag{1}[cc][cc]{$\scriptstyle 1$}
\psfrag{2}[cc][cc]{$\scriptstyle 2$}
\psfrag{3}[cc][cc]{$\scriptstyle 3$}
\psfrag{4}[cc][cc]{$\scriptstyle 4$}
\psfrag{Nc}[cc][cc]{$\scriptstyle N_c^{\sum k_i}$}
\centerline{ \includegraphics[height=35mm]{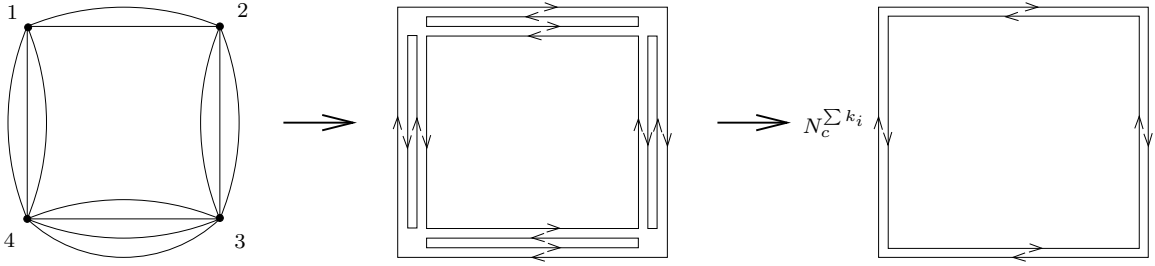} }
 \caption{\small In the planar limit each propagator is replaced by a pair of arrowed lines and the vertex is
replaced as shown in Fig.~\ref{Vertex}. Each line with an arrow corresponds to a Wilson line.}
 \label{Multi}
 \end{figure}%
\begin{figure}[h!]
\psfrag{1}[cc][cc]{$\scriptstyle 1$}
\psfrag{2}[cc][cc]{$\scriptstyle 2$}
\psfrag{3}[cc][cc]{$\scriptstyle 3$}
\psfrag{4}[cc][cc]{$\scriptstyle 4$}
\psfrag{Nc}[cc][cc]{$\scriptstyle N_c^{\sum k_i}$}
\centerline{ \includegraphics[height=35mm]{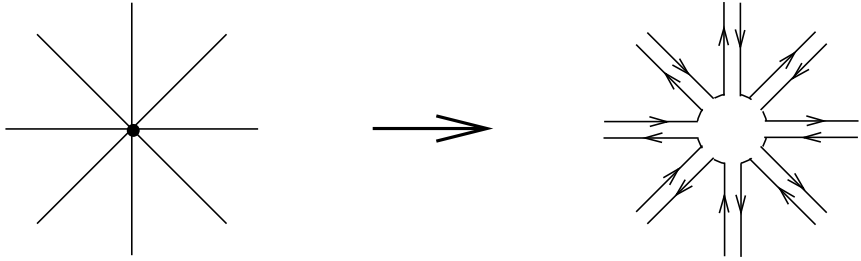} }
 \caption{\small Each vertex $q^k$ in the planar limit is replaced by $k$ pairs of arrowed lines.}
\label{Vertex}
 \end{figure}%
The diagrammatic derivation of the above relation is shown in Figure~\ref{Multi}.
As explained in the parallel paper \cite{AEKMS}, each propagator connecting a pair of adjacent points $x_i$ and $x_{i+1}$ is approximated by a free propagator multiplied by a Wilson line along the segment $[x_i , x_{i+1} ]$, evaluated in the adjoint representation, $W_{\rm adj} [x_i , x_{i+1} ]$. Then, the vertex at point $x_i$ contains $k_i$ Wilson lines with their color indices contracted to ensure that the total color charge is zero. The Wilson lines in the adjoint and in the fundamental representations are related to each other as
\begin{align}
  (W_{\rm adj} [x_i , x_{i+1} ])_{ab} t^b
= W_{\rm fund} [x_i , x_{i+1} ] t^a W_{\rm fund} [x_{i+1} , x_i ]\,,
\end{align}
or equivalently (for the gauge group $U(N)$)
\begin{align}
(W_{\rm fund} [x_i , x_{i+1} ])_{ij} (W_{\rm fund} [x_{i+1} , x_i ])_{kl} = (t^a
)_{kj} (W_{\rm adj} [x_i , x_{i+1} ])_{ab} (t^b)_{il}\,.
\end{align}
In the multi-color limit, we can use the last identity to replace a Wilson line in the adjoint representation by a pair of two parallel fundamental Wilson lines with opposite orientations. This is shown in the middle panel of Fig.~\ref{Multi}. In this way, we obtain a collection of closed cycles. We observe that all cycles but two have a backtrack shape, i.e. the corresponding contour encircles a zero area. We  denote such contour by
$C \cup C^{-1}$. Only two cycles go through all cusp points $x_i$ with different orientations. Notice that the Wilson lines satisfy the unitarity condition
\begin{align}
 W_C (W_C )^\dagger = W_C W_{C ^{-1}} = 1\,.
\end{align}
As a consequence, each backtrack cycle reduces to 1 and we arrive at the right-hand side panel in Fig.~\ref{Multi}. It contains only two cycles, each corresponding to a Wilson loop in the fundamental representation.


\end{document}